\documentstyle[12pt,epsf]{article}

\epsfverbosetrue
\newcommand{\ewdueup}[2]{\setlength{\epsfxsize}{#2}\epsfbox[10 30 640 300]{#1}}
\newcommand{\ewxy}[2]{\setlength{\epsfxsize}{#2}\epsfbox[10 30 640 590]{#1}}

\newcommand{\plus}{\makebox[15pt][c]{$+$}}
\newcommand{\minus}{\makebox[15pt][c]{$-$}}
\newcommand{\ds}{\displaystyle}
\newcommand{\be}{\begin{equation}}
\newcommand{\nn}{\nonumber}
\newcommand{\ee}{\end{equation}}
\newcommand{\bea}{\begin{eqnarray}}
\newcommand{\ba}{\begin{array}}
\newcommand{\ea}{\end{array}}
\newcommand{\eea}{\end{eqnarray}}
\newcommand{\errr}[2]{\raisebox{0.08em}{\scriptsize {$\;\begin{array}{@{}l@{}}
                          \plus\makebox[0.9em][r]{#1} \\[-0.12em] 
                          \minus\makebox[0.9em][r]{#2} 
                        \end{array}$}}}

\newcommand{\err}[2]{\raisebox{0.08em}{\scriptsize {$\;\begin{array}{@{}l@{}}
                          \plus\makebox[0.55em][r]{#1} \\[-0.12em] 
                          \minus\makebox[0.55em][r]{#2} 
                        \end{array}$}}}
\newcommand{\er}[2]{\raisebox{0.08em}{\scriptsize {$\;\begin{array}{@{}l@{}}
                          \plus\makebox[0.15em][r]{#1} \\[-0.12em] 
                          \minus\makebox[0.15em][r]{#2} 
                        \end{array}$}}}

\newcommand{\one}{1\:\!\!\!{\rm{I}}}              
\newcommand{\pslash}{\rlap{p}{\kern0.1em\hbox{/}}}
\newcommand{\qslash}{\rlap{q}{\kern0.1em\hbox{/}}}

\begin{document}
\begin{titlepage}

\begin{flushleft}
Draft, 
\today
\end{flushleft}

\begin{flushright}
Edinburgh Preprint: 95/555 \\
Granada Preprint UG-DFM-3/96 \\
Marseille CPT-95/PE.3244 \\
Southampton Preprint SHEP-95/31 \\ 
\end{flushright}

\vspace*{5mm}

\begin{center}
{\Huge Heavy Baryon Spectroscopy }\\[10mm]
{\Huge from the Lattice}\\[7mm]
{\large\it UKQCD Collaboration}\\[3mm]

{\bf K.C.~Bowler, R.D.~Kenway, O.~Oliveira, D.G.Richards,
P.~Ueberholz\footnote{Present address: Department of Physics, University
of Wuppertal, Wuppertal D-42097, Germany}}\\ 
Department of Physics \& Astronomy, The University of
Edinburgh, Edinburgh EH9~3JZ, Scotland

{\bf L.~Lellouch 
\footnote{Permanent  address: Centre de Physique Th\'eorique, CNRS Luminy, 
Case 907, F-13288 Marseille 
Cedex 9, France.}, J.Nieves\footnote{Permanent address:
Departamento de Fisica Moderna, Universidad de Granada, 18071, Spain.}, 
C.T.~Sachrajda\footnote{Address from 1 Oct. 1995 to 1 Oct. 1996: 
Theory Division, CERN, 1211
Geneva 23, Switzerland.}, N.~Stella, 
H.~Wittig\footnote{Present address: DESY-IFH, Platanenallee~6,
D-15738 Zeuthen, Germany}}\\
Physics Department, The University, Southampton SO17~1BJ, UK

\end{center}
\setcounter{footnote}{0}
\begin{abstract}

The results of an exploratory lattice study of heavy
baryon  spectroscopy are presented. We have computed the full spectrum
of the eight baryons containing  a single heavy quark, on a $24^3\times
48$ lattice at $\beta=6.2$, using an $O(a)$-improved fermion action.
We discuss the lattice baryon operators and give a method for
isolating the contributions of the spin doublets $(\Sigma,\Sigma^*)$,
$(\Xi',\Xi^*)$ and $(\Omega,\Omega^*)$ to the correlation function of
the relevant operator.  We compare our results with the available
experimental data and find good agreement in both the charm and the
beauty sectors, despite the long extrapolation  in the heavy quark mass
needed in the latter case. We also predict the masses of several
undiscovered baryons.  We compute the $\Lambda-\mbox{pseudoscalar meson}$ and
$\Sigma-\Lambda$  mass splittings. Our results, which have errors in the range $
10-30\%$, are in good agreement
with the  experimental numbers.  For the $\Sigma^*-\Sigma$ mass splitting,
we find results considerably smaller than the experimental values for both the 
charm and the beauty baryons, although in the latter case the experimental 
results are still preliminary. This is also the case for the lattice results for 
the hyperfine splitting for the heavy mesons.

\end{abstract}

\end{titlepage}

\newpage
\section{Introduction}

The discovery of the $\Lambda_b$ baryon at LEP \cite{LEP1} and the
claims of indirect evidence for $\Lambda_b$ and $ \Xi_b$ semileptonic
decays \cite{LEPSEMIL}, have triggered an increased interest
in the spectroscopy and weak decays of heavy baryons. The
interest in the spectroscopy, in particular, has been considerably
boosted after the announcement of the discovery  of 
several spin-$\frac{3}{2}$ charm and beauty baryons \cite{Peer1,Peer2}. 

The properties of hadrons containing a heavy quark can be studied
using lattice QCD calculations, which provide non-perturbative,
model-independent results. Experience gained through studies of heavy
mesons has provided the framework for an investigation of the
phenomenology of heavy baryons. Furthermore, the study of the spectrum
of heavy baryons is a necessary precondition for the measurement of
the weak matrix elements of semileptonic decays of beauty baryons. The
computed masses and matrix elements can then be combined with an analysis
carried out in Heavy Quark Effective Theory
(HQET)~\cite{NEUBERT,DESY} to extract an independent measurement of
the CKM matrix elements $V_{cb}$ and $V_{ub}$.

The subject of spectroscopy has been widely discussed in the
literature, mainly using potential models~\cite{POTENTIAL},
HQET~\cite{rosner}, or a combination of the latter with chiral
perturbation theory~\cite{CHIRAL}. Recently, there have been attempts
to compute the mass of the $\Lambda_h$ (one heavy quark and two light
quarks) \cite{lat94} and of the $\Xi_{hh}$ (two heavy quarks and one
light quark) on the lattice \cite{WUPPERTAL}.  In this paper, for the
first time, the full spectrum of the lowest-lying baryons containing one heavy
quark is computed.  In particular, we define operators suitable for
the simulation of baryon spin doublets with total spin $\frac{1}{2}$
and $\frac{3}{2}$, like the $(\Sigma,\Sigma^*)$, $(\Xi',\Xi^*)$ and
$(\Omega,\Omega^*)$.  The quality of the signal we have observed and
the agreement of our estimates with the available experimental data
are good, thus giving us confidence in the reliability of our
predictions.

The quark content and quantum numbers of the baryons we have
considered are summarised in Table \ref{tab1}. On the lattices
available at present, it is not possible to simulate directly the
beauty quark, whose mass is larger than the cutoff.  Therefore we have
computed four heavy quark masses around that of the charm quark and
interpolated (extrapolated) the results to the charm (beauty) 
quark, relying on the predictions of HQET. The masses of the charm and 
beauty quark were fixed from the masses of the $D$ and $B$ 
mesons, respectively.
The results of the extrapolation in the heavy quark mass, could be
checked carefully, given the relatively large sample of masses
available.  On the other hand, we have only used two different values
of the light quark mass, thus limiting our ability to perform a
detailed analysis of the chiral behaviour.  Our results for the masses
are given in Table~\ref{tab9}, where the first set of errors is purely
statistical and the second set is an estimate of the systematic
uncertainty in the calibration of the lattice spacing.  Our results
are in good agreement with the experimental determinations, where
available\footnote{The experimental evidence for the $\Xi'_c$ baryon
is based on a collection of 11 events, and no estimate of the
statistical error is given, see ref.~\cite{LAST1}.  We note that the
physical $\Xi'_h$ and $\Xi_h$ states are mixtures of the states which
we measure here, where the light-quark system has definite spin.  It
has been argued that such mixing \cite{Rosner1} becomes negligible in
the heavy quark limit \cite{rosner,CHIRAL}. See the conclusions for
further comments.}. We also present estimates for the $\Lambda -\mbox{
pseudoscalar meson}$ mass splittings. Our results are
\begin{equation}
\frac{M_{\Lambda_c}-M_D}{M_{\Lambda_c}+M_D}= 0.099\err{9}{7}\ \ \ \ \ 
\frac{M_{\Lambda_b}-M_B}{M_{\Lambda_b}+M_B}= 0.033\err{5}{4}
\label{ratio1}
\end{equation} 
to be compared with the experimental values\footnote{
Errors on the experimental data are added in quadrature.}   
\begin{equation}
\frac{M_{\Lambda_c}-M_D}{M_{\Lambda_c}+M_D} =  0.100(3)\ \ \ \ \
\frac{M_{\Lambda_b}-M_B}{M_{\Lambda_b}+M_B} =  0.033(5). 
\label{ratio2}
\end{equation}
Similarly, for
the $\Sigma-\Lambda$ splitting, we find 
\begin{equation}
\frac{M_{\Sigma_c}-M_{\Lambda_c}}{M_{\Sigma_c}+M_{\Lambda_c}}=0.039\err{9}{9}
\ \ \ \ \
\frac{M_{\Sigma_b}-M_{\Lambda_b}}{M_{\Sigma_b}+M_{\Lambda_b}}=0.017\err{5}{7}
\label{ratio3}
\end{equation}  
which compares well with the experimental numbers
\begin{equation}
\frac{M_{\Sigma_c}-M_{\Lambda_c}}{M_{\Sigma_c}+M_{\Lambda_c}} =
0.035(1)\ \ \ \ \ 
\frac{M_{\Sigma_b}-M_{\Lambda_b}}{M_{\Sigma_b}+M_{\Lambda_b}} =
0.016(2). 
\label{eq4}
\end{equation}
The last number in Eq.~(\ref{eq4}), extracted from the data presented in
ref.~\cite{Peer2}, is still preliminary.

We also make a first attempt to estimate the spin splitting of the
doublets $(\Sigma^*,\Sigma)$, $(\Xi^*,\Xi')$ and $(\Omega^ *,\Omega)$
by isolating the contributions which the two particles give to the
same correlation function. We find small, negative splittings, which,
in most cases, become compatible with zero after the extrapolations
because of the increased statistical errors.  The simple quark model
expectation is that the splittings are positive, although some of the
experimental data are still inconclusive. If this expectation is
confirmed by experiment, we could be facing a situation similar to
that of the hyperfine splitting in heavy meson systems, where the
splitting is underestimated using both the standard Wilson
action and the Sheikholeslami-Wohlert (SW)
action~\cite{MSCLOVER}.  The meson hyperfine splitting is sensitive to
the chromomagnetic moment term which appears at $O(a)$ in improved
fermion actions~\cite{FERMILAB}.  We plan to investigate the
sensitivity of the baryon hyperfine splitting by using the
tadpole-improved Sheikholeslami-Wohlert action~\cite{LAST2}.

\begin{table}
\begin{center}
\begin{tabular}{||c|rl|rl|c|c||}
\hline\hline
Baryon &$(S)$&$J^{P} $&$(I)$&${s_{l}}^{\pi_{l}}$& Quark Content & Operator\\
\hline 
$\Lambda_{c,b} $&$(0)$&${\frac{1}{2}}^{+} $&$(0)$&$0^{+} $& 
$(ud)c,b$&${\cal O}_5$\\
\hline
$\Sigma_{c,b} $&$(0)$&${\frac{1}{2}}^{+} $&$(1)$&$1^{+} $& $(uu)c,b $
&$\cal{O}_{\mu}$ \\
\hline
$\Sigma^*_{c,b} $&$(0)$&${\frac{3}{2}}^{+} $&$(1)$&$1^{+} $&$(uu)c,b $
&$\cal{O}_{\mu}$ \\
 \hline
$\Xi_{c,b} $&$(-1)$&${\frac{1}{2}}^{+} $&$(\frac{1}{2})$&$0^{+} $
&$(us)c,b$&${\cal O}_5$
\\ \hline
$\Xi'_{c,b} $&$(-1)$&${\frac{1}{2}}^{+} $&$(\frac{1}{2})$&$1^{+} $&$(us)c,b $
&$\cal{O}'_{\mu}$ \\
 \hline
$\Xi^*_{c,b} $&$(-1)$&${\frac{3}{2}}^{+} $&$(\frac{1}{2})$&$1^{+} $& $(us)c,b$
&$\cal{O}'_{\mu}$  
 \\ \hline
$\Omega_{c,b} $&$(-2)$&${\frac{1}{2}}^{+} $&$(0)$&$1^{+} $&$(ss)c,b$
&$\cal{O}_{\mu}$ 
\\  \hline
$\Omega^*_{c,b} $&$(-2)$&${\frac{3}{2}}^{+} $&$(0)$&$1^{+} $&$(ss)c,b $
&$\cal{O}_{\mu}$ 
\\ \hline\hline
\end{tabular}
\end{center}
\caption{\em Summary of the quantum numbers of the eight 
baryons containing a single heavy quark.   $I$,
$s_l^{\pi_l}$ are the isospin and the spin-parity of the light degrees
of freedom and $S$, $J^P$ are the strangeness and the spin-parity of
the baryon.} \label{tab1} %
\end{table}

This paper is organised as follows: in Section 2 we discuss the
baryonic operators  which have been used in the present study and give
details of the simulation. In Section 3 we explain our analysis
procedures for the  extraction of the masses. The measurement of the
mass splittings  is reported in Section 4. Our results and the
comparison with the physical values are reported  in Section 5.
Finally,  we present our conclusions.

\section{Particles and Operators}

There are eight lowest-lying 
baryons containing one heavy and two light quarks (up,
down or  strange). The  quantum numbers of the charm and beauty
baryons  are listed in Table~\ref{tab1}, and their 
physical masses (see refs.
\cite{Peer1,Peer2,PDB,SKAT}), 
are given in Table~\ref{tab9}. In the context of HQET at
lowest order, it is possible to identify the spin-parity quantum
numbers of the heavy quark and of the light system, within each
baryon. Furthermore, heavy baryons with common light degrees of
freedom exhibit common features; they are expected to be degenerate in
mass, and to obey selection rules in weak decays. For example, the
hyperfine mass splittings in the doublets $(\Sigma^*,\Sigma)$,
$(\Xi^*,\Xi')$ and $(\Omega^*,\Omega)$ are expected to be $O(1/m_h)$, where 
$m_h$ is the mass of the heavy quark,
and the weak semileptonic decay $\Lambda_b\to\Sigma_c$ is suppressed
since it could only take place if the light quark system changed
quantum numbers.

\begin{table}[t]
\begin{center}
\begin{tabular}{||cc|cc|cc||}\hline\hline
           &          & $\ \ h= $ & $\ charm $ & $\ \ h= $ & $\ beauty $ \\
Baryon& Quark &   Exp.      & Latt.           &  Exp.       
& Latt.            \\
          & Content &$[MeV]$ &$[GeV]$       &$[MeV]$   &$[GeV]$         \\   
\hline 
$\Lambda_{h}   $& $(ud)h$& 2285(1) &$ 2.27\er{4}{3}\er{3}{3}$& 5641(50) 
&$ 5.64\er{5}{5}\er{3}{2} $ \\
$\Sigma_{h}    $& $(uu)h$ &  2453(1)&$ 2.46\er{7}{3}\er{5}{5}$& 5814(60)
& $ 5.77\er{6}{6}\er{4}{4} $ \\
$\Sigma^*_{h}  $& $(uu)h$& 2530(7)&$ 2.44\er{6}{4}\er{4}{5}$& 5870(60)  
& $ 5.78\er{5}{6}\er{4}{3} $\\
$\Xi_{h}       $& $(us)h$&2468(4)&$ 2.41\er{3}{3}\er{4}{4}$&               
& $ 5.76\er{3}{5}\er{4}{3} $\\
$\Xi'_{h}      $& $(us)h$     & $2560^{\dagger}\ \ $ 
&$ 2.57\er{6}{3}\er{6}{6}$&               & $ 5.90\er{6}{6}\er{4}{4} $\\
$\Xi^*_{h}     $& $(us)h$     & 2643(2)   &$ 2.55\er{5}{4}\er{6}{5}$
&               & $ 5.90\er{4}{6}\er{4}{5} $\\
$\Omega_{h}    $& $(ss)h$ & 2704(20)&$ 2.68\er{5}{4}\er{5}{6}$
&             & $ 5.99\er{5}{5}\er{5}{5} $\\
$\Omega^*_{h}  $& $(ss)h$&               &$  2.66\er{5}{3}\er{6}{7}$
&               & $ 6.00\er{4}{5}\er{5}{5} $  \\   
\hline\hline
\end{tabular}
\end{center}
\caption{\em Heavy baryons considered in this project. 
Our results are quoted with a statistical error (first) and a
systematic error (second) arising from the uncertainty in the
calibration of the lattice spacing.  Where available, we report the
experimental data.\newline $(^{\dagger}$ For the error on the
$\Xi'_c$ mass, see footnote 1.) }
\label{tab9}
\end{table}

\subsection{Operators for Heavy Baryons}
\label{ops}

The spectrum of the heavy baryons in Table \ref{tab1} can be computed 
on the lattice by using the following three interpolating operators:
\begin{equation}
{\cal O}_5=\epsilon_{abc} (l^{aT} {\cal C} \gamma_5 l'^b)h^c;\ \ \
{\cal O_{\mu}}=\epsilon_{abc} (l^{aT} {\cal C} \gamma_{\mu} l^b)h^c; \ \ 
{\cal O'_{\mu}}=\epsilon_{abc} (l^{aT} {\cal C} \gamma_{\mu} l'^b)h^c
\label{operators}
\end{equation}
where $a,b,c$ are colour indices, $\epsilon_{abc}$ is the totally
antisymmetric Levi-Civita tensor, ${\cal C}$ is the charge conjugation
matrix, $l,l'$ are light quark fields, and $h$ is the heavy-quark
field. The (implicit) spinorial index of the three operators, 
is the (implicit) uncontracted Dirac index carried by the heavy quark field. By
employing the operators in Eq.~(\ref{operators}), one creates physical
states whose heavy quark and light quark systems have definite quantum
numbers, corresponding to the HQET description at lowest order.

The operator ${\cal O}_5$ corresponds to $s_l^{\pi_l} = 0^+$
spin-parity for the light degrees of freedom  and a total spin-parity
for the baryon $J^P=\frac{1}{2}^+$. The total isospin of the light
degrees of freedom is $I=0$ if $l=u$ and $l'=d$, and $I=\frac{1}{2}$ if
 one of the light quarks is the strange quark.

Consider the  two-point correlation function $G_5$:
\begin{equation}
G_5(\vec{p},t)=
\sum_{\vec{x}} e^{-i\vec{p}\cdot\vec{x}}\langle{\cal O}_5(\vec{x},t) 
\overline{{\cal O}}_5(\vec{0},0)\rangle.
\end{equation}
For large time separations, using antiperiodic boundary conditions in
the time direction, this becomes in the continuum limit
\begin{eqnarray}
G_5(\vec{p},t) &\to& 
\frac{Z^2}{2E} \big\{ e^{-Et} (M+ \pslash) -e^{-E(T-t)}(M-\tilde{\pslash}) 
\big\} \nonumber \\
&+&\frac{Z^2_{PP}}{2E_{PP}} \big\{ e^{-E_{PP}t} (M_{PP}-\qslash ) -
e^{-E_{PP}(T-t)}(\tilde{\qslash} + M_{PP}) \big\} 
\label{total5}
\end{eqnarray}
where $p^{\mu}=(E,\vec{p})$ is the 4-momentum of the baryon and 
$\tilde{p}^{\mu} = (E,-\vec{p})$. In Eq.~(\ref{total5}),  we have
included the contribution of the parity partner (PP) baryon, with
4-momentum $q^{\mu}=(E_{PP},\vec{p})$ and  $\tilde{q}^{\mu} =
(E_{PP},-\vec{p})$. This particle contributes to the correlation
function because it has  a non-zero overlap with the operator 
${\cal O}_5$ given in Eq.~(\ref{operators}).
At zero momentum, we choose an appropriate
combination of spinorial indices such that the baryon, but not the
parity partner, propagates forward in time. We find that the amplitude
of the parity partner ($Z^2_{PP}$) propagating backward in time is 
much smaller than that of the forward-propagating 
baryon ($Z^2$), and, in the following,
we will neglect the contribution of the parity partners.

The case of the operator ${\cal O}_{\mu}$ is more involved than that
of ${\cal O}_5$ since it transforms reducibly under the parity
extended Lorentz group. It is the tensor product of a
four vector and a Dirac spinor and thus transforms as
$(1,\frac{1}{2})\oplus (\frac{1}{2},1)$$\ 
\oplus\ (\frac{1}{2},0)\oplus
(0,\frac{1}{2})$ (in $SU(2)\otimes SU(2)$ notation).  It can
annihilate/create particles of spin-parity $\frac{3}{2}^+$ and
$\frac{1}{2}^+$ as well as these particles' parity partners. 
With the two-point function for the operator ${\cal O_{\mu}}$ defined as
\begin{equation}
G_{\mu\nu}(\vec{p},t)=
\sum_{\vec{x}} e^{-i\vec{p}\cdot\vec{x}}<{\cal O}_{\mu}(\vec{x},t) 
\overline{{\cal O}}_{\nu}(\vec{0},0)>,
\label{opemunu}
\end{equation}
we find, in the continuum limit,  
for large values of $t$ and using antiperiodic boundary 
conditions in time, 
\begin{eqnarray}
G_{\mu\nu}(\vec{p},t) \to &&\frac{Z_{3/2}^2}{2E_{3/2}} 
e^{-E_{3/2}t} (\pslash_{3/2} + M_{3/2})(P^{3/2})_{\mu\nu}(p_{3/2})
\nonumber\\
&&+ \frac{e^{-E_{1/2}t}}{2E_{1/2}} 
\Big\{Z_1^2 (\pslash_{1/2} + M_{1/2})(P^{1/2}_{11})_{\mu\nu}(p_{1/2})
-Z_2^2 (\pslash_{1/2} - M_{1/2})(P^{1/2}_{22})_{\mu\nu}(p_{1/2})
\nonumber\\
&&-Z_1Z_2 (\pslash_{1/2} + M_{1/2})(P^{1/2}_{12})_{\mu\nu}(p_{1/2})
+Z_2Z_1 (\pslash_{1/2} - M_{1/2})(P^{1/2}_{21})_{\mu\nu}(p_{1/2})\Big\}
\nonumber\\
&&+ \mbox{parity partners} - \mbox{antiparticles} \ ,
\label{totalMN}
\end{eqnarray}
where $p_J^\mu=(E_J,\vec{p})$ and where parity partner contributions
are obtained from the original particle contributions with the
replacement $M_J\to -M_{PP,J}$ while antiparticle contributions are
obtained from the original particle and parity partner contributions
with the replacement $M\to -M$, $\vec p\to -\vec{p}$ and $t\to T-t$ in
the exponent. For any given momentum $p_{\mu}$, the quantities
$(P^{3/2})_{\mu\nu}(p)$ and $(P^{1/2}_{ij})_{\mu\nu}(p)$, $i,j=1,2$,
are the spin projection operators of ref.~\cite{BEN} and are given by
\begin{eqnarray}
(P^{3/2})_{\mu\nu}(p) &=&
g_{\mu\nu} - \frac{1}{3} \gamma_{\mu}\gamma_{\nu} - \frac{1}{3p^2}\left(
\pslash\gamma_{\mu}p_{\nu} + p_{\mu}\gamma_{\nu}\pslash\right) 
\ ,\nonumber\\
(P^{1/2}_{11})_{\mu\nu}(p) &=&
 \frac{1}{3} \gamma_{\mu}\gamma_{\nu} -\frac{1}{p^2}p_{\mu}p_{\nu} 
+ \frac{1}{3p^2}\left(
\pslash\gamma_{\mu}p_{\nu} + p_{\mu}\gamma_{\nu}\pslash\right)
\ ,\nonumber\\
(P^{1/2}_{22})_{\mu\nu}(p)&=&\frac{1}{p^2}p_{\mu}p_{\nu}
\ ,\\
(P^{1/2}_{12})_{\mu\nu}(p)&=&\frac{1}{\sqrt{3}p^2}\left(p_{\mu}p_{\nu}
-\pslash\gamma_\mu p_\nu\right)
\nonumber\\
(P^{1/2}_{21})_{\mu\nu}(p)&=&\frac{1}{\sqrt{3}p^2}\left(\pslash p_\mu 
\gamma_\nu-p_{\mu}p_{\nu}\right)
\ .\nonumber
\label{projectors}
\end{eqnarray}
They are orthonormal and idempotent
\begin{equation}
(P^{I}_{ij})_{\mu\rho}
(P^{J}_{kl})^{\rho\nu}
= \delta^{IJ}\delta_{jk}(P^{J}_{il})_{\mu}^{\nu}
\label{ortho}
\end{equation}
where $I,J$ take on values
$\frac{1}{2}$ or $\frac{3}{2}$. They satisfy
$$
\gamma_{\mu} (P^{3/2})^{\mu}_{\nu}= 0\ ;
\qquad\qquad p_{\mu} (P^{3/2})^{\mu\nu}= 0 = (P^{3/2})^{\mu\nu} p_\nu\ ;
$$
\begin{equation}
p_{\mu} (P^{1/2}_{1j})^{\mu\nu}= 0 = (P^{1/2}_{i1})^{\mu\nu} p_\nu
\quad\mbox{for $i,j=1,2$}
\label{rw1}
\end{equation}
and have the following useful properties
\begin{eqnarray}
\pslash (P^{1/2}_{ij})^{\mu\nu}&=& 
\pm (P^{1/2}_{ij})^{\mu\nu} \pslash\quad\mbox{$+$ for $i=j$, $-$ for $i\ne j$}
\nonumber\\
\pslash (P^{3/2})^{\mu\nu}&=& 
(P^{3/2})^{\mu\nu} \pslash .
\label{rw}
\end{eqnarray}
The properties of Eqs.~(\ref{rw1}) and (\ref{rw}) 
guarantee that the spin-$\frac{3}{2}$ contribution
satisfies the appropriate Rarita-Schwinger equations~\cite{RS} and
that the spin-$\frac{1}{2}$ contributions satisfy the appropriate Dirac
equations.  The diagonal projectors are furthermore complete:
\begin{equation}
g_{\mu\nu}=(P^{3/2})_{\mu\nu}(p) + (P^{1/2}_{11})_{\mu\nu}(p) + 
(P^{1/2}_{22})_{\mu\nu}(p)
\ .
\end{equation}

To extract the masses 
of the spin-parity $\frac{1}{2}^+$ and
$\frac{3}{2}^+$ particles, one needs only to compute the correlators
(\ref{totalMN}) at rest. In this case, the projectors $P^{3/2}$
and $P^{1/2}_{11}$ simplify to
\begin{eqnarray}
(P^{3/2})^{ij}& =& g^{ij}-\frac{1}{3}\gamma^i\gamma^j ; \ \ 
i,j=1,2,3  , \nonumber\\
(P^{1/2}_{11})^{ij}& =& \frac{1}{3}\gamma^i\gamma^j ,
\label{rfproj}
\end{eqnarray}
and only act on the spatial components of $G_{\mu\nu}(\vec{0},t)$,
i.e.  $\mu,\ \nu=1,2,3$.  Since the components corresponding to the
other projection operators do not contribute to the spatial
components, $G_{ij}(\vec{0},t)$, it is clear from the properties of
$P^{3/2}$ and $P^{1/2}_{11}$ given in Eqs.~(\ref{ortho}) and
(\ref{rw}) that the $\frac{3}{2}^+$ contribution can be isolated by
considering $(P^{3/2})^{ij}G_{jk}(\vec{0},t)$ and the $\frac{1}{2}^+$
contribution, by considering $(P^{1/2}_{11})^{ij}G_{jk}(\vec{0},t)$.
The contributions of forward-propagating parity partners are
suppressed by taking suitable combinations of spinorial indices as
discussed after Eq.~(\ref{total5}) for the case of $G_5(\vec{0},t)$.
Those of the backward-propagating parity partners are naturally
smaller because the time intervals over which they propagate are much
longer for most values of $t$ that we consider in analysing
$G_{jk}(\vec{0},t)$.  Furthermore, both contributions are again
empirically found to be suppressed by the fact that the overlaps of
the parity partner states with the operator ${\cal O}^\mu$ are orders
of magnitude smaller and their masses slightly larger than those of
the original particles.

When space-time is approximated by a hypercubic lattice, full
Euclidean $O(4)$ symmetry is reduced to symmetry under the hypercubic group.
This reduction means that most irreducible representations of $O(4)$
and its covering group become reducible on the lattice. Fortunately,
the representations which concern us here, $(1,\frac{1}{2})$,
$(\frac{1}{2},1)$, $(\frac{1}{2},0)$ and $(0,\frac{1}{2})$, because of
their low dimensionality, remain irreducible on the
lattice~\cite{MANDULA}. Furthermore, when restricted to the diagonal
cubic subgroup (i.e. the lattice analogue of the rotation subgroup),
$(1,\frac{1}{2})\oplus (\frac{1}{2},1)$ decomposes 
into the reducible representation
$\frac{3}{2}\oplus\frac{1}{2}$ while $(\frac{1}{2},0)\oplus (0,\frac{1}{2})$
reduces to $\frac{1}{2}$, where $\frac{1}{2}$ and $\frac{3}{2}$ are 
themselves reductions to the cubic group of continuum 
spin-$\frac{1}{2}$ and spin-$\frac{3}{2}$ representations. Thus, the
space-time transformation properties of the operators ${\cal O}_5$
and ${\cal O}_{\mu}$ on the lattice are analogous to
what they are in the continuum. Moreover, using the results
of Ref.~\cite{MANDULA}, one can show that the cubic representations
$\frac{1}{2}$ and $\frac{3}{2}$ mix only with the following continuum
spins:
\begin{equation}
\left(\frac{1}{2}\right)_{\mbox{cubic}}:
\quad \frac{1}{2},\frac{7}{2},\frac{9}{2},\ldots
\quad\quad\left(\frac{3}{2}\right)_{\mbox{cubic}}:
\quad \frac{3}{2},\frac{5}{2},
\frac{7}{2},\ldots
\ .
\label{spinmix}
\end{equation}
Therefore, if one isolates correctly the cubic $\frac{1}{2}$ and
$\frac{3}{2}$ contributions to $G_{\mu\nu}(\vec{p},t)$, one isolates
unambiguously the contributions of the continuum spin-$\frac{1}{2}$
and spin-$\frac{3}{2}$ states in the large time limit (assuming, of
course, that higher spin states are more massive).  It should be
emphasized that this isolation of the cubic representations must be
done carefully because it is not known, a priori, which of the
spin-$\frac{1}{2}$ or the spin-$\frac{3}{2}$ states is more massive.
Fortunately, at zero momentum the continuum rest frame projectors given
in Eq.~(\ref{rfproj}) are sufficient because they implement the
Clebsch-Gordan decomposition of the product representation $\mbox{spin-}1
\otimes \mbox{spin-}\frac{1}{2}$ into spin-$\frac{1}{2}$ and
spin-$\frac{3}{2}$, a decomposition which survives the reduction of
$SU(2)$ to the double valued cubic subgroup because irreducible
representations of $SU(2)$ with spin less than or equal to $\frac{3}{2}$
are irreducible when reduced to that subgroup. It should further be
noted that properties of operators and states under parity
tranformations are unaffected by the discretization of space-time.

A similar discussion applies to the operator ${\cal O}'_{\mu}$.  In
fact, the structure of the corresponding correlator in terms of quark
propagators is the same as that of ${\cal O}_\mu$; the only effect of
the additional Wick contraction in the ${\cal O}_\mu$ correlator is to
change the overall normalization.

Thus we have shown in the present section how to isolate the contributions
of the different physical baryon states to the two-point functions of
Eqs.~(\ref{total5}) and (\ref{totalMN}). In the following sections we
shall use these procedures to determine the heavy-light baryon
spectrum. To improve the overlap of the interpolating operators ${\cal
O}_5$, ${\cal O}_{\mu}$ and ${\cal O}'_{\mu}$ with the corresponding
physical baryon states, we smear these operators as described in the
Appendix.  Though this smearing further complicates the isolation of
the various physical baryon state contributions for non-zero two-point
function momentum $\vec{p}$ (please refer to the Appendix for
details), the baryon spectrum that we obtain here
only requires an analysis of zero-momentum, smeared two-point
functions to which the discussion of the present section applies
unchanged.

\subsection{Details of the Simulation}

Our calculation is performed on 60 $SU(3)$ gauge field configurations
generated on a $24^3\times48$ lattice at $\beta=6.2$, using the hybrid
over-relaxed algorithm described in ref.\,\cite{a_units}. The quark
propagators were computed using the $O(a)$-improved
Sheikholeslami-Wohlert action, which is related to the standard
Wilson fermion action via
\begin{equation}
S_F^{SW} = S_F^W - i\frac{\kappa}{2}\sum_{x,\mu,\nu}\bar{q}(x)
         F_{\mu\nu}(x)\sigma_{\mu\nu}q(x),
\end{equation}
where $\kappa$ is the hopping parameter. 
The use of the SW action reduces discretisation errors from
$O(ma)$ to $O(\alpha_s\,ma)$  \cite{SW,HEATLIE}, which is of particular
importance in our study of heavy baryons, where the bare heavy quark
masses are typically around one third to two thirds of the inverse
lattice spacing. 

The gauge field configurations and light quark propagators were
generated on the 64-node i860 Meiko Computing Surface at the University
of Edinburgh.  The heavy quark propagators were computed using the
256-node Cray T3D, also at Edinburgh.

Statistical errors are calculated according to the bootstrap procedure described
in \cite{a_units}, for which the quoted errors on all quantities correspond
to 68\% confidence limits of the distribution obtained from 1000 bootstrap
samples.

In order to convert our values for baryon masses and mass splittings into 
physical units we need an estimate of the inverse lattice spacing in GeV.
In this study we take
\begin{equation}
\label{eq:ainv}
a^{-1} = 2.9 \pm 0.2\,{\rm \ GeV},
\end{equation}
thus deviating slightly from some of our earlier papers, where we quoted
2.7 GeV  as the central value \cite{a_units,strange,fd}.
The error in equation (\ref{eq:ainv}) is large enough to encompass all
our estimates for $a^{-1}$ from quantities such as $m_\rho$, $f_\pi$,
$m_N$, the  string tension $\sqrt{K}$ and the hadronic scale $R_0$ 
discussed in
\cite{Sommer}. This change is partly motivated by a recent study using
newly generated UKQCD data \cite{biele_how}; using the quantity $R_0$
we found $a^{-1} = 2.95\err{7}{11}$ GeV. Also, the non-perturbative
determination of the renormalisation constant of the axial current
yielded a value of $Z_A = 1.05(1)$ \cite{jonivar} which is larger by
about 8\% than the perturbative value which we used previously. Thus
the scale estimated from $f_\pi$ decreases to around 3.1 GeV which
enables us to reduce significantly the upper uncertainty on our final
value of $a^{-1}$ [GeV].

Light quark propagators were computed for quark masses around the
strange quark mass, corresponding to hopping parameters $\kappa=$
0.14144 and 0.14226.  Because each heavy baryon contains two light
quarks, we can form three baryon correlators for each heavy quark
mass, of which two have degenerate light-quark masses and one has
non-degenerate light-quark masses.  The masses of the light
pseudoscalar meson which are needed for this study, were obtained in
ref.~\cite{strange}.  Results extrapolated to the chiral limit
(corresponding to a hopping parameter $\kappa_{\rm crit} =
0.14315\er{2}{2}$\,) and to the mass of the strange quark
($\kappa_{\rm s} = 0.1419\er{1}{1}$\,) are also tabulated there.

The heavy quark propagators have been computed for four values of the
heavy quark mass around that of the charm quark, corresponding to $\kappa_h=$
0.133, 0.129, 0.125 and 0.121. The masses of the heavy-light
pseudoscalar mesons can be found in ref. \cite{fd}. 

In order to enhance the signal for the baryon correlation functions,
the light and heavy quark propagators have been computed using the
Jacobi smearing method \cite{SMEARING}, either at the source only (SL)
or at both the source and the sink (SS). Since smearing is not a
Lorentz-invariant operation, it might alter some of the transformation
properties of non-scalar observables. We have found that such an
effect is evident in the baryonic correlators at non-zero momentum,
and we present the results of a study of these effects in
Appendix~\ref{app2}. This issue, which does not affect the spectrum,
represents an important new effect which is crucial in the extraction
of the amplitude $Z$, and therefore in the measurement of the weak
matrix elements entering the semileptonic decays of the $\Lambda_b$.

\section{Analysis Details}

It follows from Eq.~(\ref{total5}), 
that, for $t>0$
\begin{equation}
[G_5(\vec{0},t)]_{11}=[G_5(\vec{0},t)]_{22}=-
[G_5(\vec{0},T-t)]_{33}=-[G_5(\vec{0},T-t)]_{44}.
\end{equation}
Therefore, we define the $\Lambda$ correlation function as
\footnote{$\Lambda$
is a conventional name, by which we mean the baryon whose light quarks
are in a $s^{\pi}=0^+$ state. Depending on the flavour of the latter,
this baryon is either the physical $\Lambda (ud)$ or the $\Xi (us)$
with spin 0 for the light quarks. A similar convention is used for
$\Sigma$ and $\Sigma^*$.}:
\begin{equation}
G_{\Lambda}(t)=\frac{1}{4}\left[
[G_5(\vec{0},t)]_{11}+G_5(\vec{0},t)]_{22}-
[G_5(\vec{0},T\!-\!t)]_{33}-G_5(\vec{0},T\!-\!t)]_{44}\right]\simeq Z_{\Lambda}^2\  e^{-m_{\Lambda}t}
\end{equation}
Similarly, for the $\Sigma$ and $\Sigma^*$, we define the correlation
functions by taking suitable combinations of the equivalent components,
after projection with the operators given in
Eq.~(\ref{rfproj}).

\subsection{The Effective Masses}

\begin{figure}[b]
\leavevmode
\begin{picture}(80,100)
\put(0,40){\ewdueup{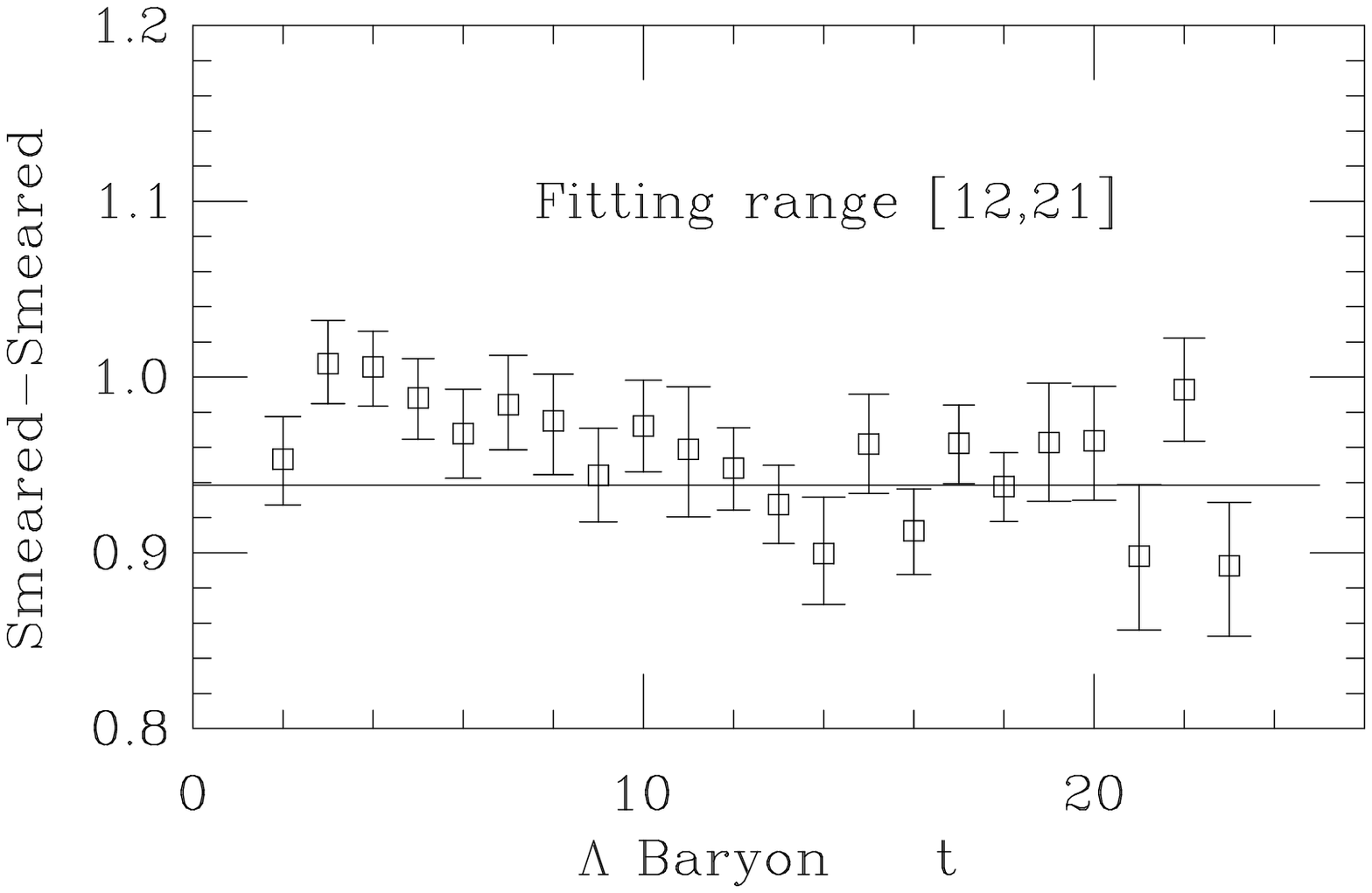}{90mm}}
\put(0,-7){\ewdueup{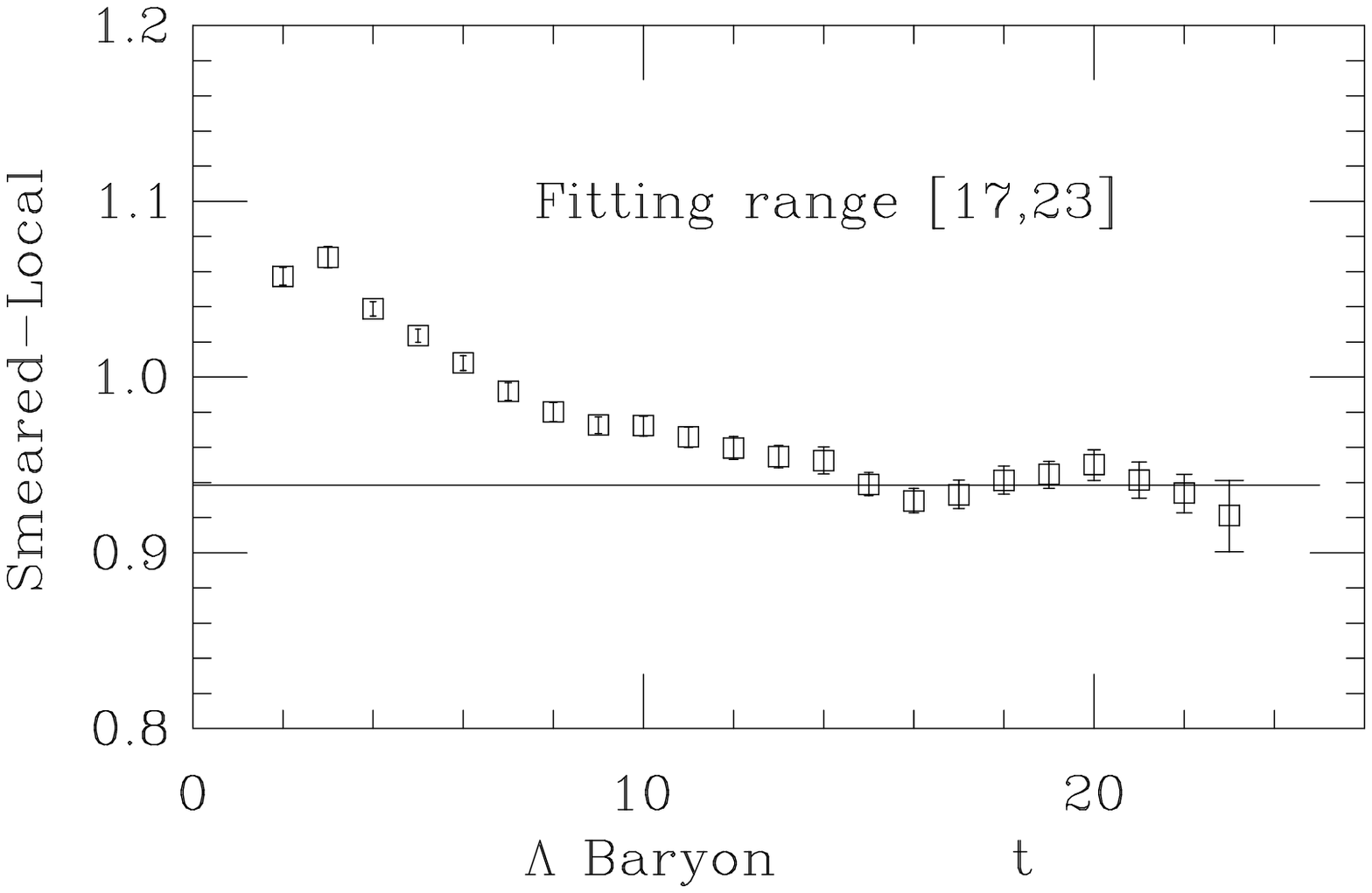}{90mm}}
\put(80,40){\ewdueup{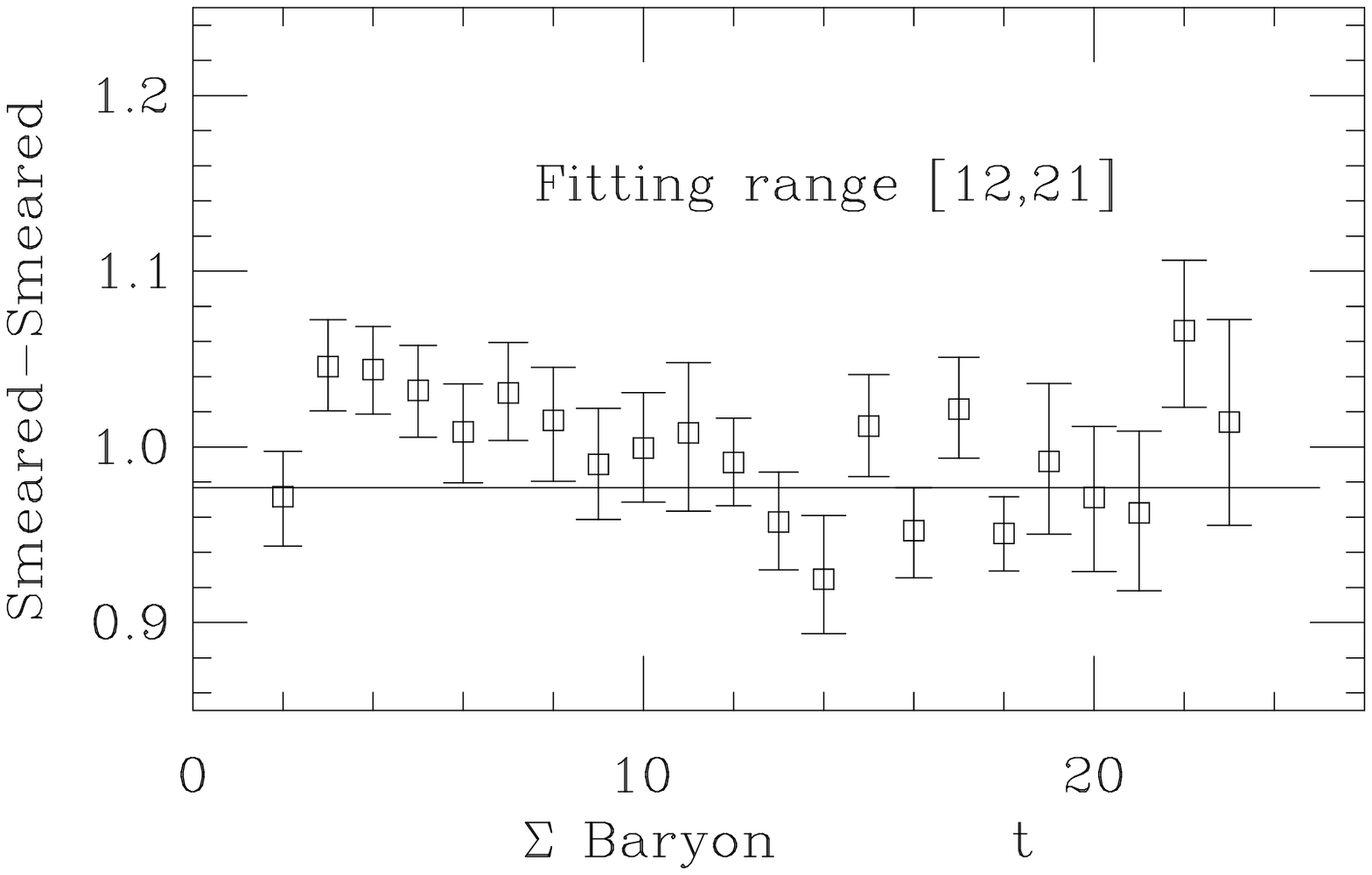}{90mm}}
\put(80,-7){\ewdueup{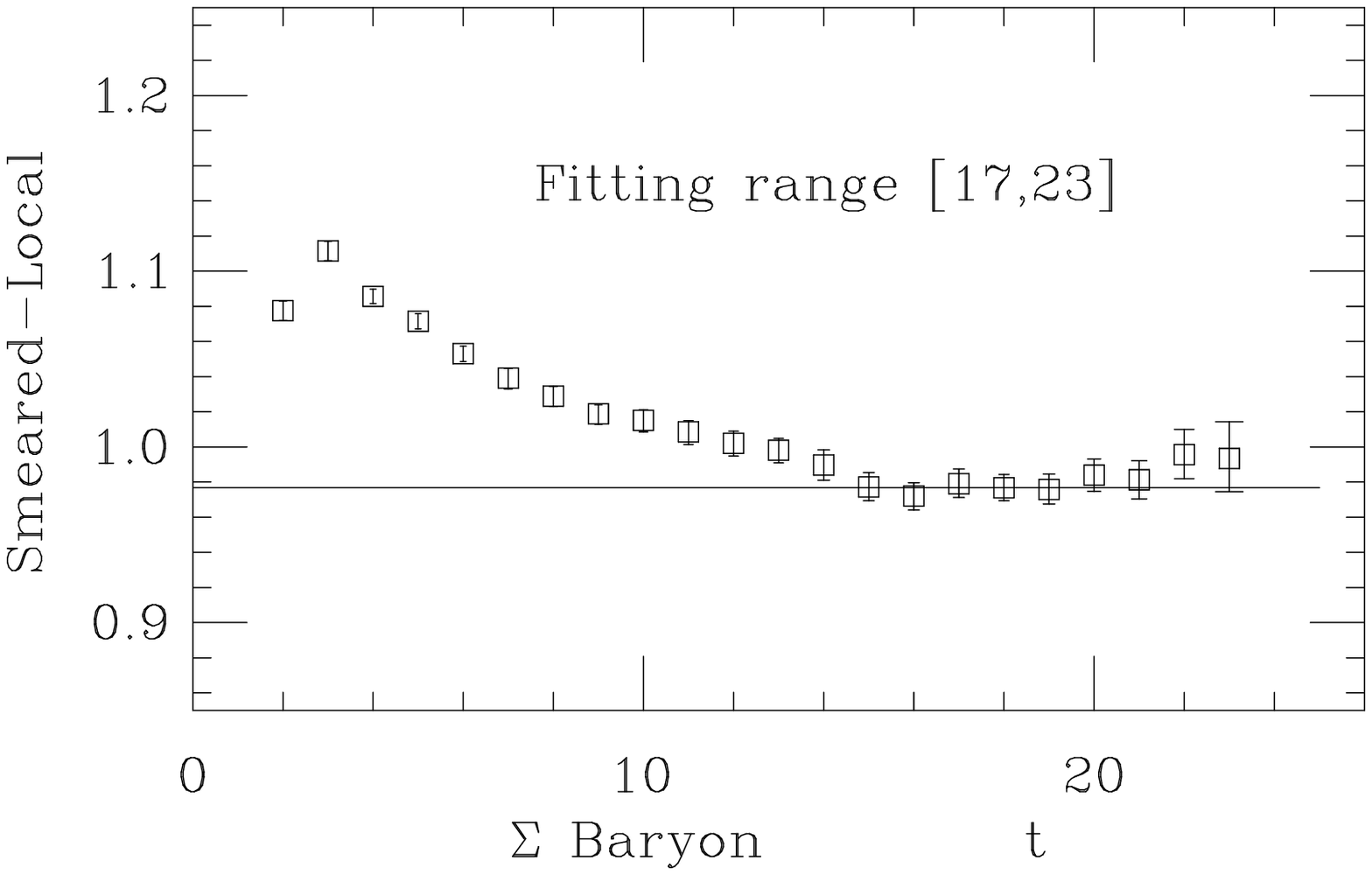}{90mm}}
\end{picture}
\caption{\em Effective masses for the $\Lambda$ and $\Sigma$ baryons. 
We show typical 
plots, corresponding to $\kappa_h=0.129$ and
$\kappa_{l1}=\kappa_{l2}=0.14144$.  
The straight lines are our best fits, which agree for SL and SS correlators
starting from $t_{\rm min}=17$ onwards.}
\label{fig1}
\end{figure}
 
In Fig.~\ref{fig1} we show effective mass plots of the $\Lambda$ and
$\Sigma$ baryons, in both the SL and SS cases. We compute the effective
masses assuming that the correlators' time evolution is given by an
exponential:
\begin{equation}
M_{\rm eff}(t) = \ln \left( \frac{G_{\Lambda,\Sigma}(t)}
{G_{\Lambda,\Sigma}(t+1)} \right).
\label{meff}
\end{equation}
This is
justified since we have checked that the contribution of the 
parity partners propagating
backward in time is completely negligible.

The effective mass is smoother for SL than for SS correlators, because
the former are more correlated in time. To establish a fitting range,
we fitted the correlators to
a single exponential in  the range $[t_{\rm min},t_{\rm max}]$, where
$t_{\rm max}$ was fixed at  21 for SS and 23 for SL correlators, and
$t_{\rm min}$ was varied  between 8 and 19. The  fits at fixed
values of the light and heavy kappas were obtained
by minimizing the $\chi^2$ computed using the full covariance matrix.

\begin{figure}[h]
\leavevmode
\begin{picture}(80,100)
\put(0,40){\ewdueup{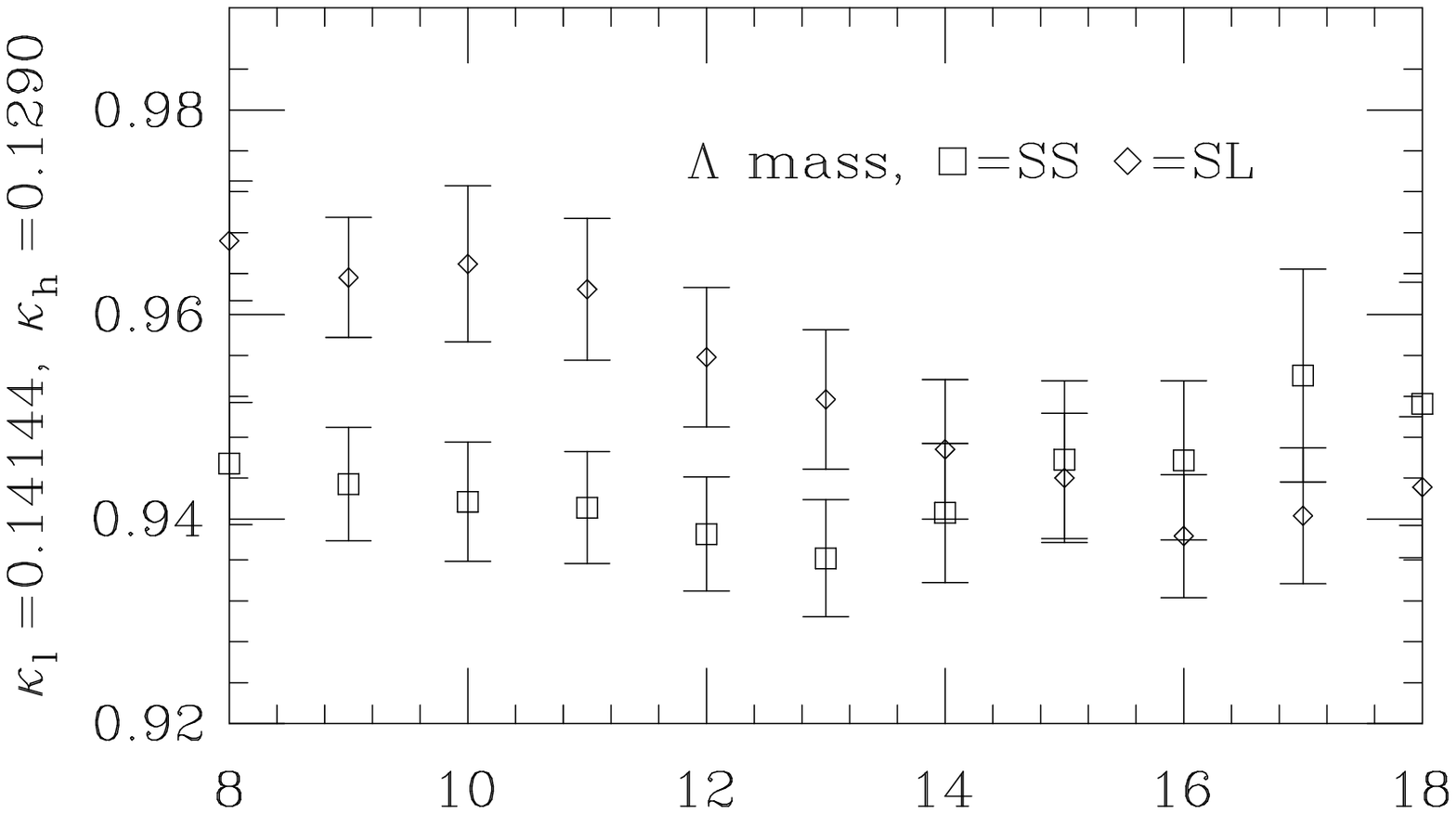}{90mm}}
\put(0,0){\ewdueup{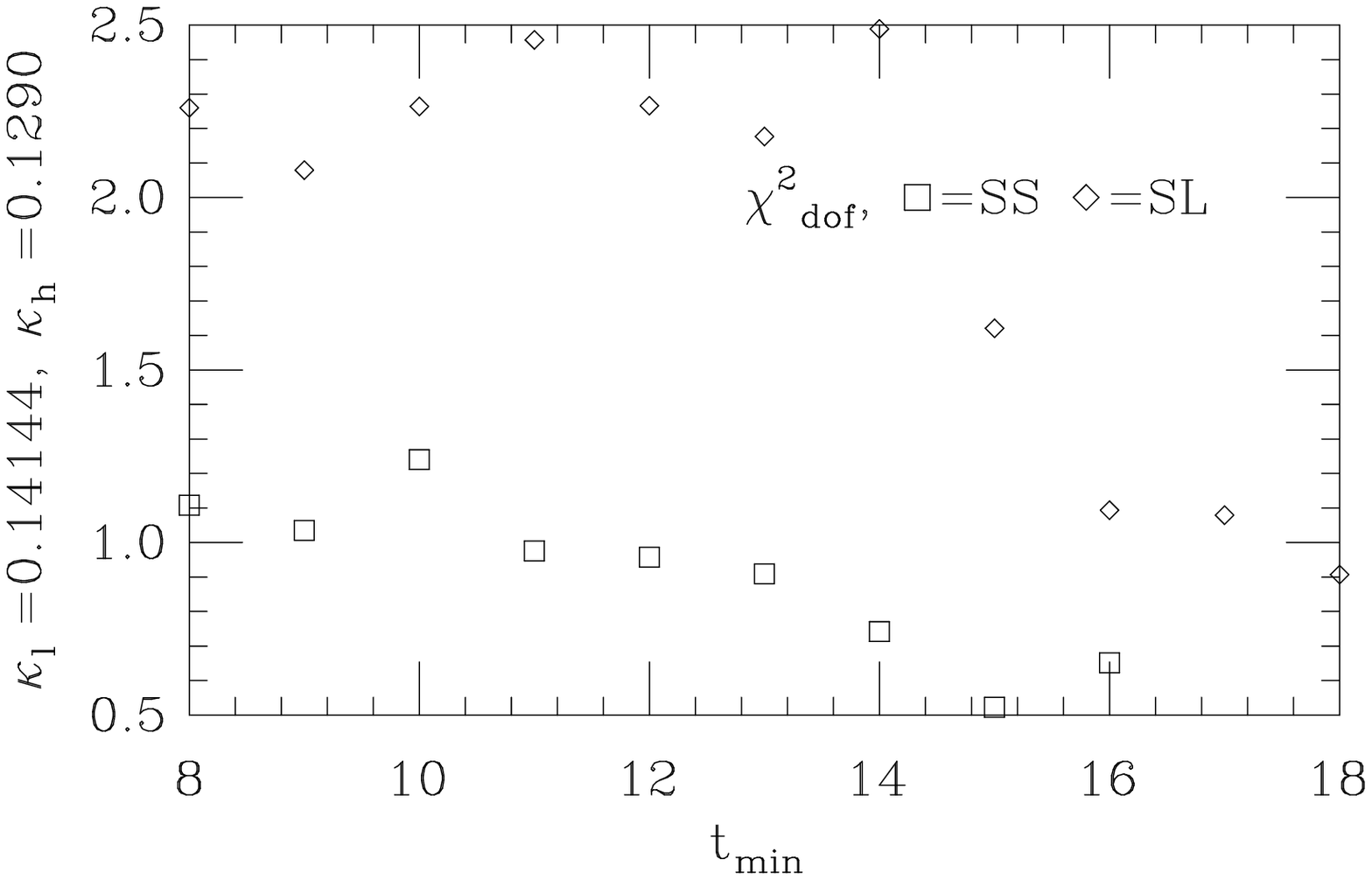}{90mm}}
\end{picture}
\leavevmode
\begin{picture}(80,100)
\put(0,40){\ewdueup{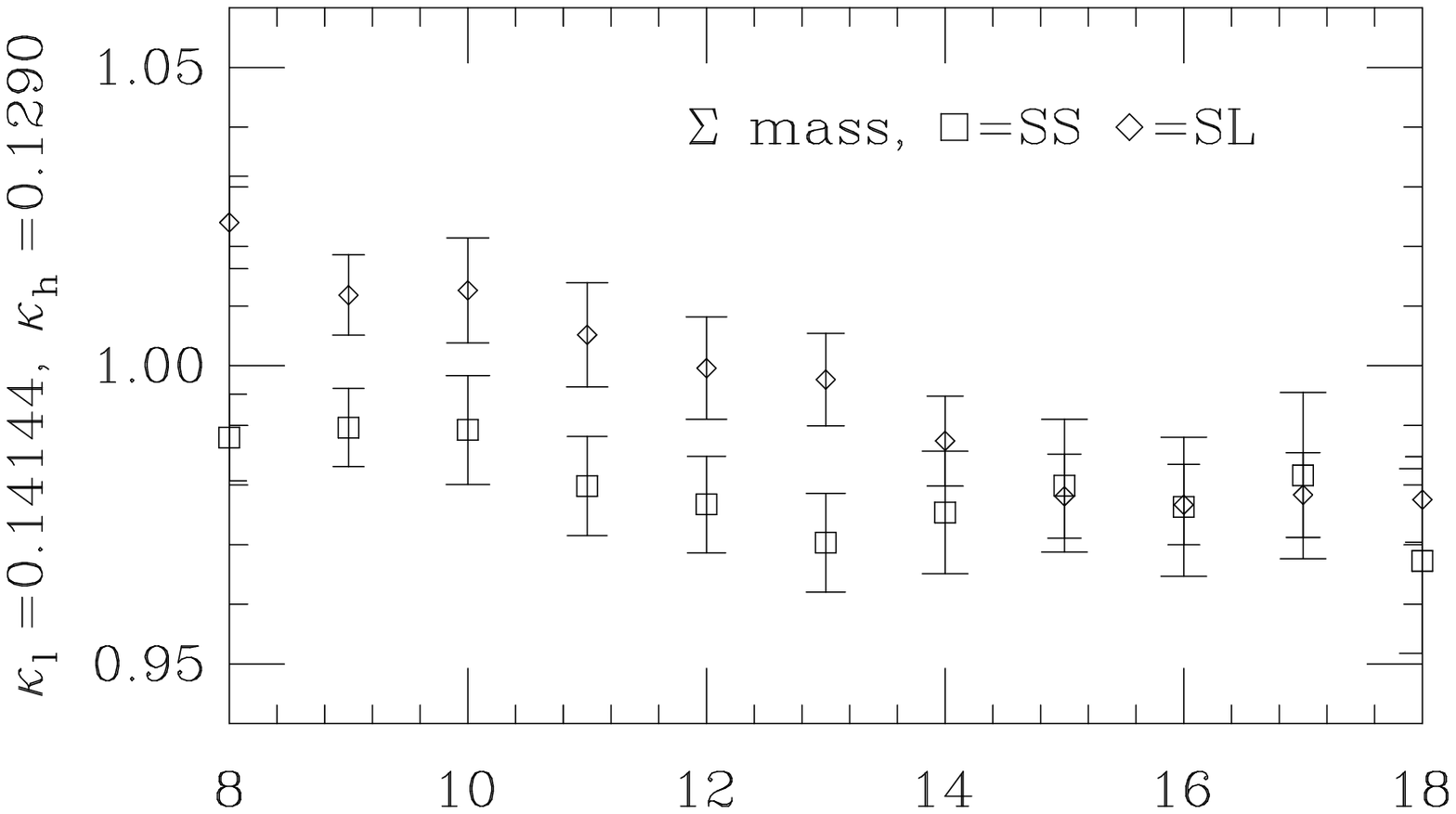}{90mm}}
\put(0,0){\ewdueup{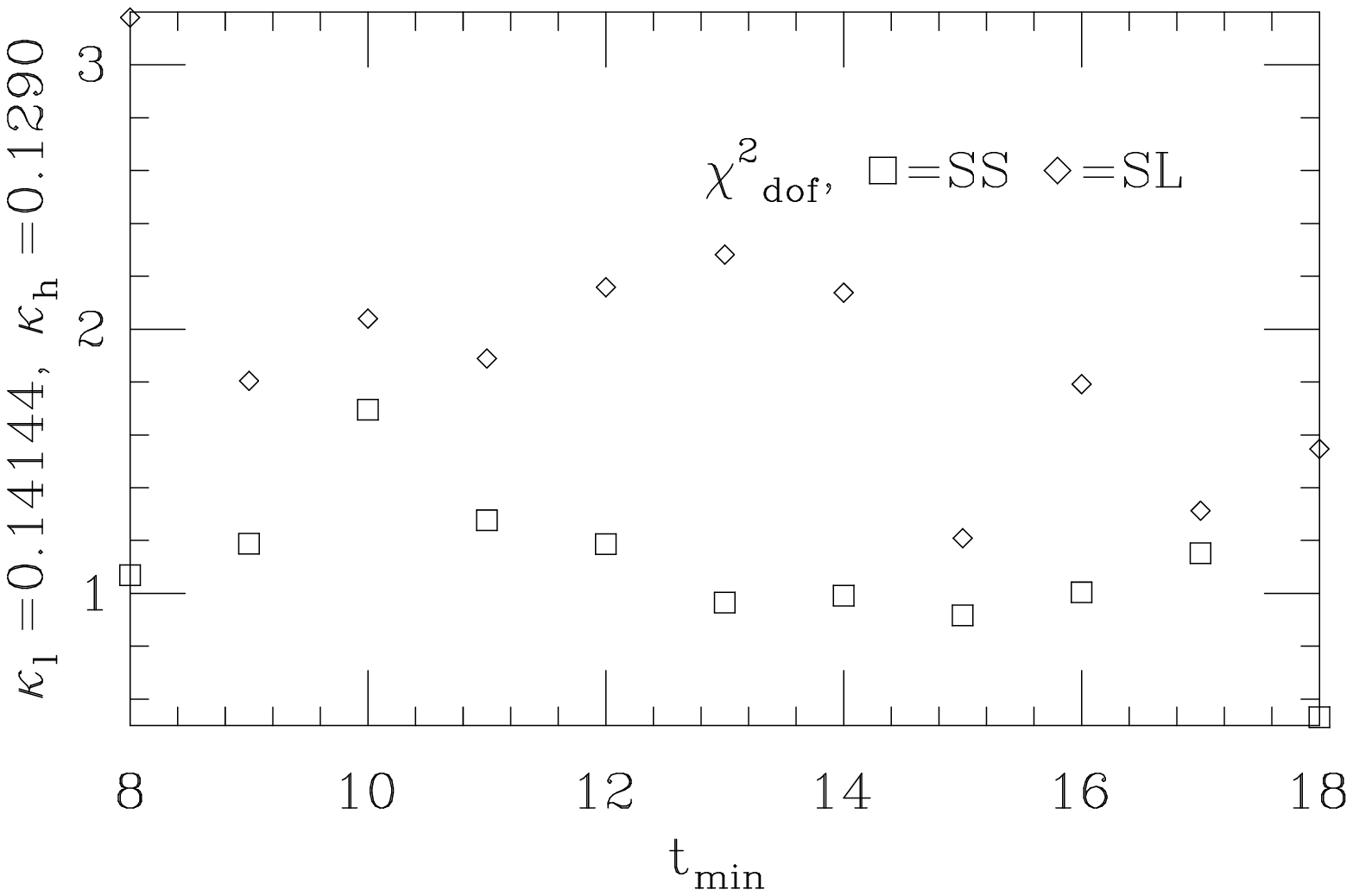}{90mm}}
\end{picture}
\caption{\em Masses and $\chi^2/{\rm dof}$ obtained from a sliding
window analysis for the $\Lambda$ and the $\Sigma$ baryon correlators.}
\label{fignew}
\end{figure}

As an example, we show in Fig.~\ref{fignew} the results of this
analysis for both the $\Lambda$ and the $\Sigma$, with
$\kappa_h=0.129$ and $\kappa_{l1}=\kappa_{\l2}=0.14144$, for both SS
and SL correlators. The behaviour of the correlator for the $\Sigma^*$
baryon is similar, and the features described below are common to all
the masses considered in this study.

By considering the $\chi^2/{\rm dof}$ of the fits, as well as the stability of
the results under variation of $t_{\rm min}$, we make the following 
observations:
\begin{enumerate} 
\item
The masses obtained from the fits to the SS correlators are stable and the
$\chi^2/{\rm dof}$ are acceptable for $t_{\rm min} \geq 11$.  For the
SL correlators, the $\chi^2/{\rm dof}$ are acceptable only for $t_{\rm
min} \geq 16$.  This behaviour supports the hypothesis that by
smearing both the sink and the source one enhances the overlap with
the ground state.
\item
The masses obtained from fits to the SL and SS correlators agree around
$t_{\rm min}\geq 17$
\item
As a general feature, we observe that the statistical errors
increase with decreasing light quark mass. This effect is more
pronounced for SL than for SS correlators.
\end{enumerate}  
The conclusion is that there is good agreement between SS and SL  data,
even if, in the  latter cases, the plateaux are shorter and the errors
slightly larger. Thus we quote the results obtained by
fitting SS correlators in $t\in[12,21]$, using those obtained with
SL correlators as a consistency check.

\subsection{Mass Extrapolations}

We obtain the masses of the eight charm and beauty baryons  by
extrapolating first in the light quark masses and then in the heavy
quark mass.

\subsubsection{Extrapolation in the Light Quark Mass}

\begin{figure}
\leavevmode
\ewxy{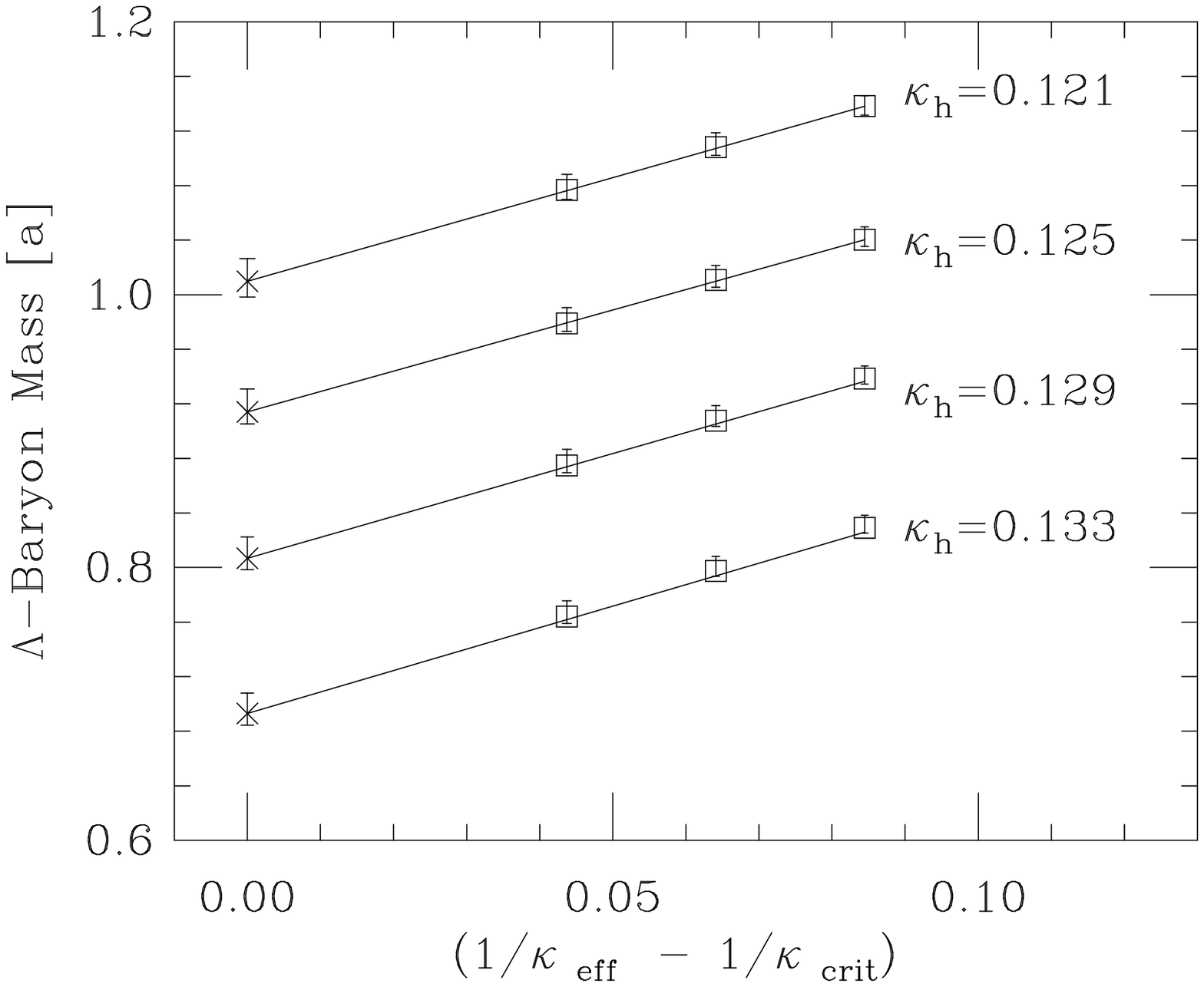}{90mm}
\leavevmode
\ewxy{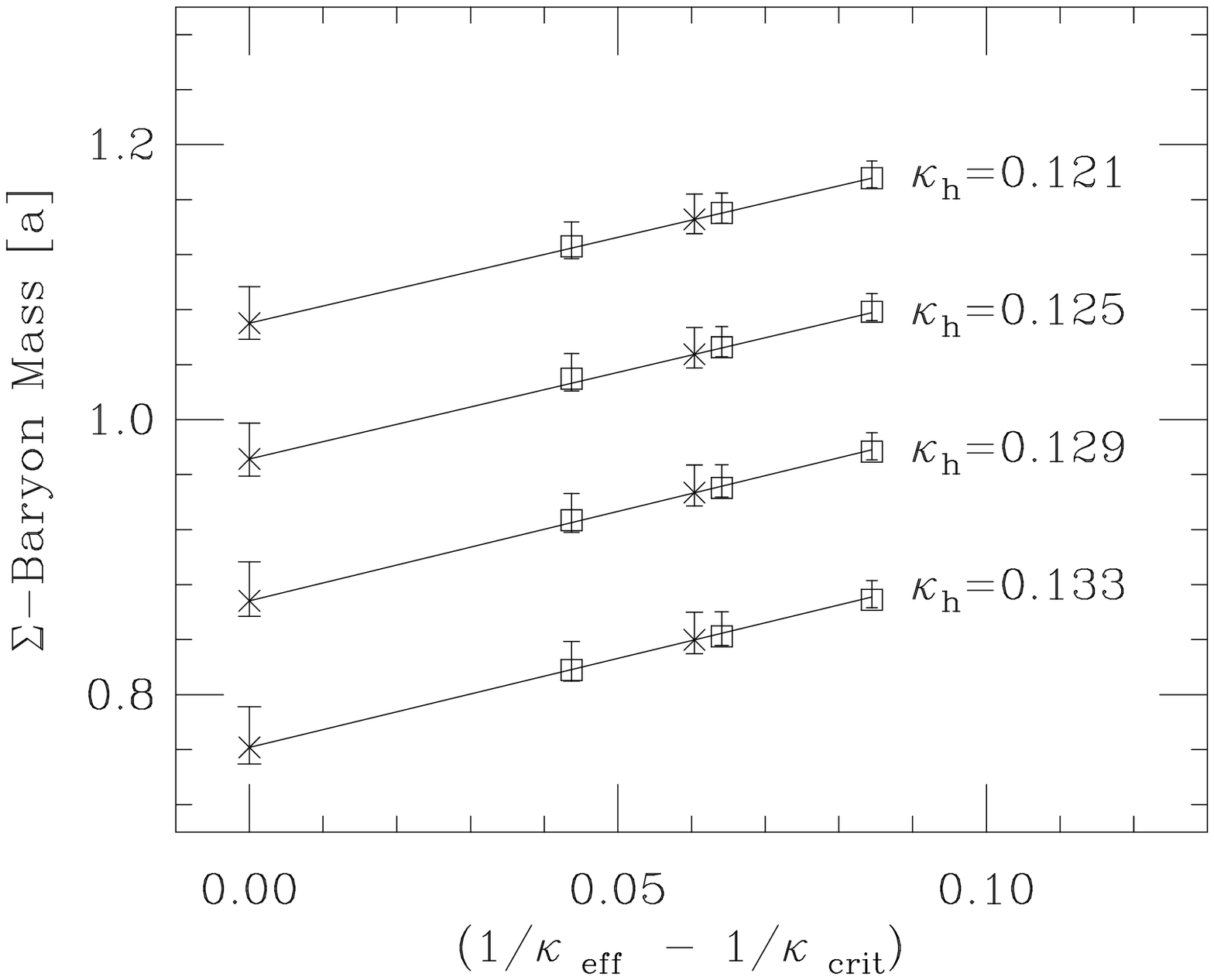}{90mm}
\caption{\em The chiral behaviour of the
$\Lambda$ and $\Sigma$ masses.
The boxes denote data at our three light-quark masses; the crosses
denote the extrapolation of our results to the chiral limit and the
strange-quark mass.}
\label{fig4}
\end{figure}

In order to perform the extrapolation to the chiral limit, we use the
three baryon masses obtained from both degenerate (i.e.\
$\kappa_{l1}=\kappa_{l2}=0.14144,$ or $0.14226$) and  non-degenerate
(i.e.\ $\kappa_{l1}=0.14144$ and $\kappa_{l2}=0.14226$) light  quark
correlation functions. We assume that, in the chiral regime, $M_{\rm
baryon}$ depends linearly  on the sum of the two light quark masses,
that is,
\begin{eqnarray}
M_{\rm baryon}(\kappa_h,\kappa_{l1},\kappa_{l2}) 
& = & M_{\rm baryon}(\kappa_h) 
      + C\left(\frac{1}{2\kappa_{l1}} + \frac{1}{2\kappa_{l2}} 
      - \frac{1}{\kappa_{\rm crit}} \right) \nonumber \\
& = & M_{\rm baryon}(\kappa_h) 
      + C\left(\frac{1}{\kappa_{\rm eff}} 
      - \frac{1}{\kappa_{\rm crit}} \right) 
\label{degenerate}
\end{eqnarray}
with $\kappa_{\rm eff}^{-1} = (\kappa_{l1}^{-1} +
\kappa_{l2}^{-1})/2$.  This is supported by our data for the masses in
the $\Lambda$ and $\Sigma$ channels, as shown in Fig.~\ref{fig4}, and
by previous studies of light meson masses~\cite{strange,gupta}. In the
$\Lambda$ channel, extrapolating to
$\kappa_{l1}=\kappa_{l2}=\kappa_{\rm crit}$ gives the mass of the
$\Lambda_h$, while extrapolating $\kappa_{l1}$ to $\kappa_{\rm crit}$
and interpolating $\kappa_{l2}$ to $\kappa_s$ gives the mass of the
$\Xi_h$.  Similarly, from the $\Sigma$ channel we extract the masses
of the $\Sigma_h$, the $\Xi'_h$ and the $\Omega_h$.

The results of this analysis, obtained from SS correlation  functions,
are summarised in Table~\ref{tab3}.  By performing the same type of
analysis on the SL correlators, we obtained essentially
indistinguishable results. We note that the difference in the
statistical errors of the SS and SL masses, already present in the fits
at fixed $\kappa$, is amplified by the extrapolation to the chiral
limit. This confirms our earlier conclusion that by using SS data one
obtaines more precise results.

\begin{table}
\begin{center}
$\Lambda\ : $ \ \ 
\begin{tabular}{||l|cccc||}
\hline\hline
$\kappa_l$     &$\kappa_h=0.121 $& $\kappa_h=0.125$ &$\kappa_h=0.129$  &$\kappa_h=0.133$     \\ \hline
0.14144/0.14144 & $  1.138\err{8}{7} $ & $  1.040\err{9}{5} $ & $ 0.939\err{9}{4} $ & $  0.829\err{9}{4} $ \\
0.14144/0.14226 & $  1.108\err{11}{6} $ & $ 1.011\err{11}{5} $ & $  0.908\err{11}{4} $ & $  0.798 \err{11}{4} $ \\
0.14226/0.14226 & $  1.077\err{11}{7} $ & $  0.979\err{12}{6} $ & $  0.875\err{12}{5} $ & $  0.764\err{12}{5} $ \\ 
\hline
strange/strange &  $  1.101\err{13}{9} $ & $  1.002\err{13}{9} $ & $  0.899\err{13}{9} $ & $  0.786\err{13}{9} $ \\
strange/chiral  &  $  1.056\err{14}{9} $ & $  0.959\err{15}{7} $ & $  0.853\err{14}{6} $ & $  0.740\err{14}{7} $ \\
chiral/chiral & $  1.011\err{17}{13} $ & $  0.913\err{16}{11} $ & $  0.807\err{15}{10} $ & $  0.693\err{14}{10} $ \\
\hline
$\ \ \ \chi^2_{\rm dof}$  & 0.05  & 0.3 &0.2 &0.9 \\
\hline
\end{tabular}   

$\Sigma\ : $ \ \ 
\begin{tabular}{||l|cccc||}
\hline
$\kappa_l$     &$\kappa_h=0.121 $ & $\kappa_h=0.125$ &$\kappa_h=0.129$  &$\kappa_h=0.133$     \\ \hline
0.14144/0.14144 & $  1.176\err{12}{7} $ & $  1.078\err{13}{6} $ & $  0.977\err{14}{6} $ & $  0.869\err{14}{6} $ \\
0.14144/0.14226 & $  1.150\err{15}{7} $ & $  1.053\err{15}{7} $ & $  0.950\err{17}{7} $ & $  0.842\err{18}{7} $ \\
0.14226/0.14226 & $  1.126\err{18}{9} $ & $  1.030\err{18}{9} $ & $  0.927\err{19}{9} $ & $  0.818\err{21}{8} $ \\ 
\hline
strange/strange &  $  1.141\err{18}{11} $ & $  1.043\err{18}{10} $ & $  0.941\err{18}{11} $ & $  0.833\err{18}{11} $ \\
strange/chiral  &  $  1.108\err{22}{10} $ & $  1.010\err{22}{11} $ & $  0.908\err{23}{10} $ & $  0.801\err{25}{10} $ \\
chiral/chiral & $  1.067\err{23}{23} $ & $ 0.965\err{23}{12} $ & $ 0.862\err{23}{10} $ & $  0.753\err{23}{13} $ \\
\hline
$\ \ \ \chi^2_{\rm dof}$  &   0.8 & 0.1 &1.0 & 0.7 \\ 
\hline
\end{tabular}   

$\Sigma^* : $ \ \ 
  \begin{tabular}{||l|cccc||}
\hline
$\kappa_l$     &$\kappa_h=0.121 $& $\kappa_h=0.125$ &$\kappa_h=0.129$  &$\kappa_h=0.133$     \\ \hline
0.14144/0.14144 & $  1.170\err{11}{7} $ & $  1.072\err{12}{7} $ & $  0.969\err{12}{6} $ & $  0.860\err{12}{6} $ \\
0.14144/0.14226 & $  1.145\err{13}{8} $ & $  1.047\err{13}{8} $ & $  0.944\err{14}{7} $ & $  0.834\err{14}{7} $ \\
0.14226/0.14226 & $  1.121\err{15}{9} $ & $  1.023\err{15}{9} $ & $  0.920\err{16}{9} $ & $  0.809\err{17}{8} $ \\ 
\hline
strange/strange &  $  1.136\err{16}{10} $ & $  1.038\err{16}{10} $ & $  0.934\err{16}{9} $ & $  0.823\err{17}{9} $ \\
strange/chiral  &  $  1.104\err{18}{11} $ & $  1.005\err{18}{10} $ & $  0.902\err{19}{10} $ & $  0.791\err{20}{9} $ \\
chiral/chiral & $  1.061\err{22}{13} $ & $  0.960\err{22}{12} $ & $  0.856\err{22}{11} $ & $  0.745\err{22}{10} $ \\
\hline
$\ \ \ \chi^2_{\rm dof}$   & 0.4 & 0.7  & 0.6 & 0.5\\
\hline\hline
\end{tabular}   
\caption{\em 
Masses of the $\Lambda$, $\Sigma$ and $\Sigma^*$ in lattice units
obtained by fitting the SS correlators in $t\in[12,21]$.  Also shown
are the corresponding masses after extrapolation in the light-quark
masses, using Eq.~\protect{\ref{degenerate}}.}
\label{tab3}
\end{center}
\end{table}

\subsection{Heavy Quark Extrapolation} 

The physical masses of the charmed and beauty baryons are obtained by
extrapolating the four sets of data, computed at $\kappa_h=0.133,
0.129, 0.125$ and $0.121$. In performing these extrapolations, we have
been guided by the HQET and have expressed the dependence of the
baryon mass $M_{\rm baryon}(M_P)$ on the heavy-light pseudoscalar
meson mass, $M_P$, through the following function 
\begin{equation}
M_{\rm baryon}(M_P) = M_P + C + \frac{A}{M_P}
\label{HS} 
\end{equation}
where the two constants $C$ and $A$ are the parameters of the fit.
The masses of the charm and beauty baryons are obtained for $M_P=M_D$ or
$M_P=M_B$ respectively.
In Table~\ref{tab9} we report the results, corresponding to the SS case,
in physical units. The numbers corresponding to the charm and beauty
masses have been obtained assuming $a^{-1}=2.9$ GeV. The quoted
systematic error arises solely from the uncertainty in the scale and
has been estimated by letting $a^{-1}$ vary by one standard deviation
about its central value.

\begin{figure}
\leavevmode
\ewxy{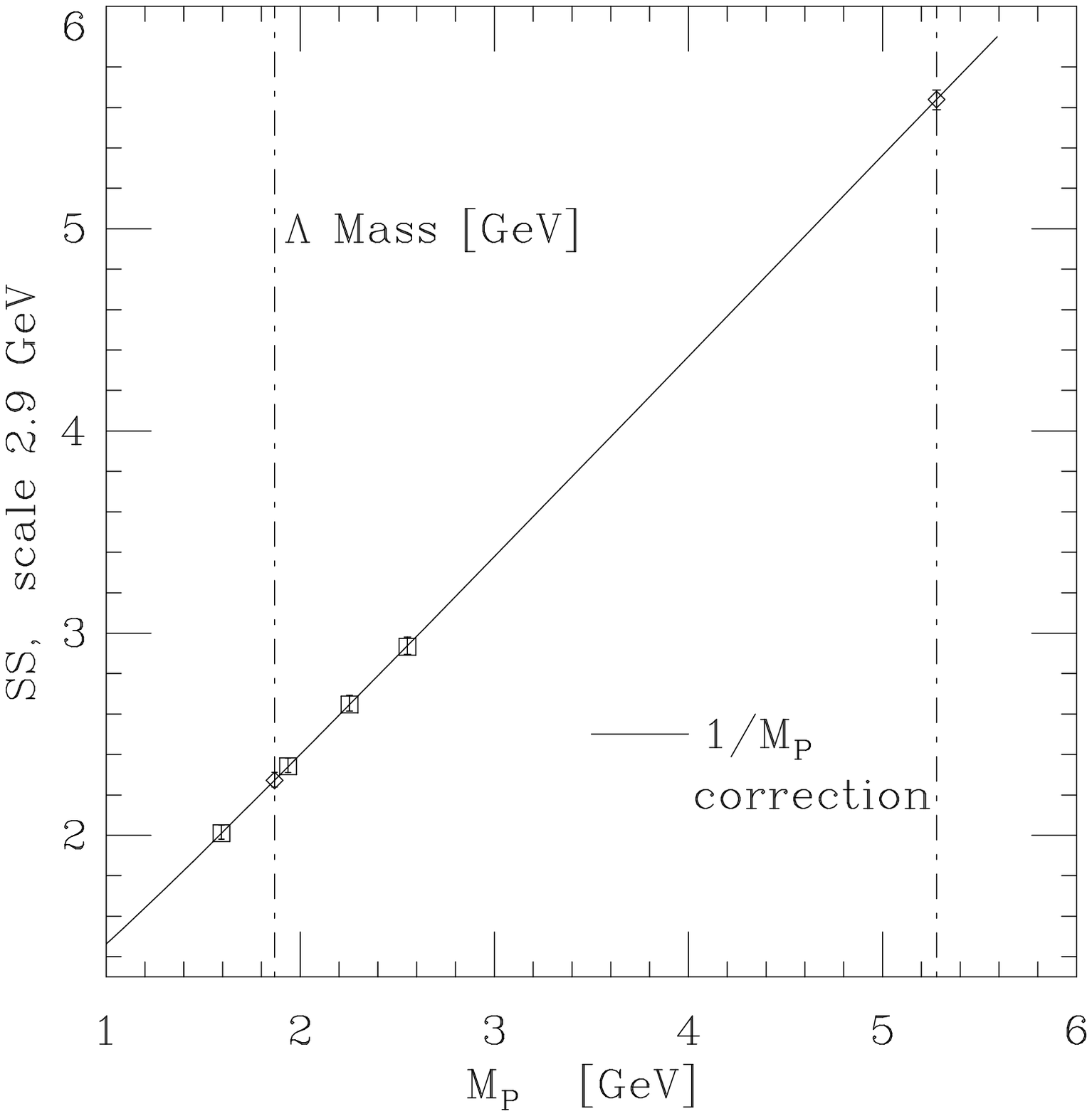}{90mm}
\leavevmode
\ewxy{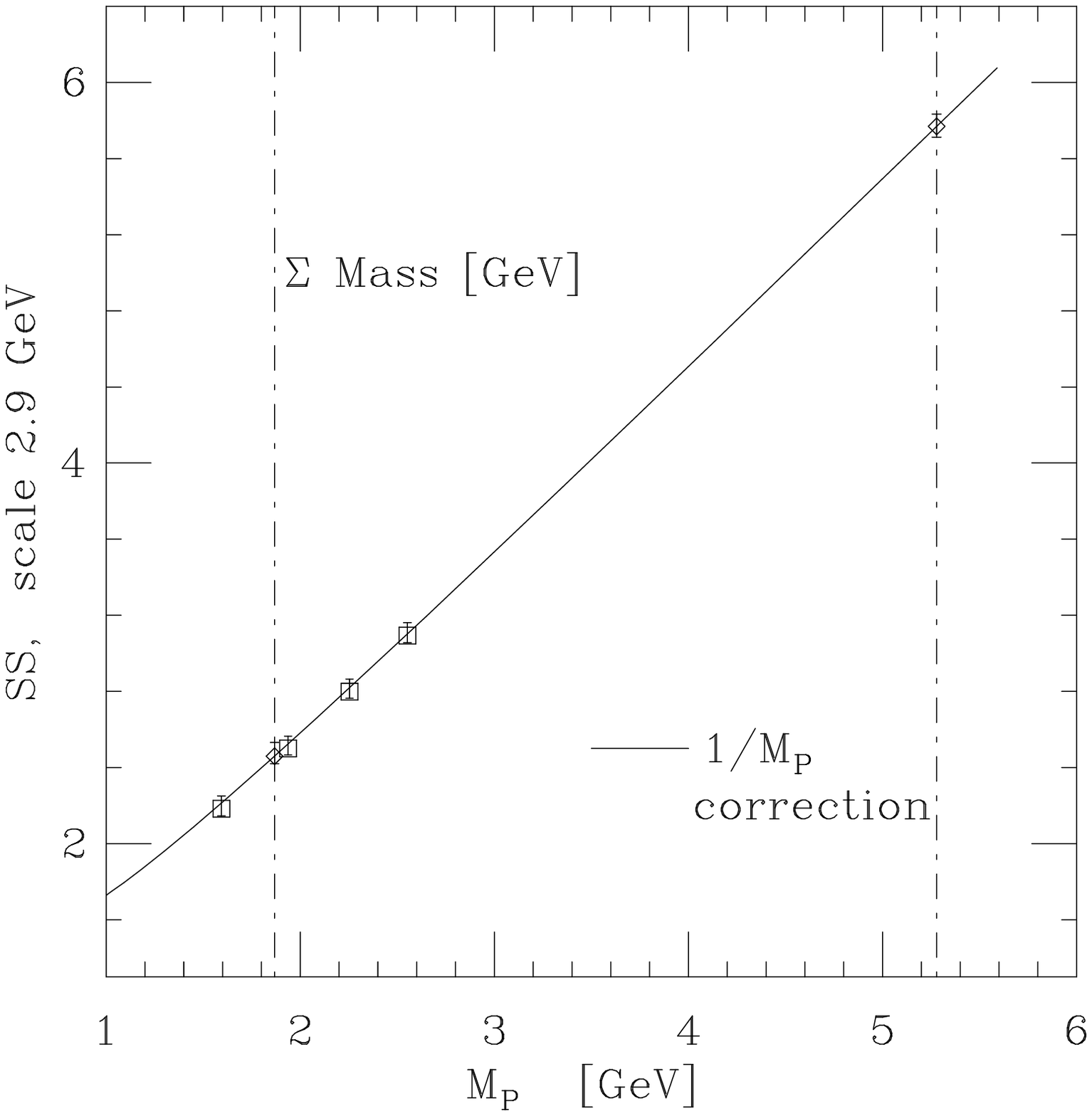}{90mm}
\caption{\em Extrapolation of the $\Lambda$ and $\Sigma$ baryon masses.
In the figures both the linear extrapolation and that obtained taking
into account $O(1/M_P)$ correction are shown. The diamonds and crosses
correspond to the extrapolations to the charm mass and beauty mass
respectively.}
\label{heavy}
\end{figure}

In these fits, the coefficient $A$, which quantifies the size of the $1/m_h$ 
corrections,
is of the expected size, i.e.
$A=O(\Lambda_{\rm QCD}^2)$, ranging from
about $(350 {\rm MeV})^2$ to $(540 {\rm MeV})^2$, depending on the particular baryon and
on the flavour of the light degrees of freedom. Of course, the 
$O(1/m_h)$ corrections play an important role in the case 
of the mass splittings,  (see section 4), while 
they contribute much less to the value of each  mass.
As a further confirmation of this, we have also set $A$ to zero and
verified that the results of the extrapolation are essentially
indistinguishable from those presented in Table~\ref{tab9} although
the $\chi^2$ are significantly higher. Finally we have used a function
of the kind 
\begin{equation} M_{\rm baryon}(M_P) = C +
A'\left(M_P -M_0\right) 
\end{equation}
where a different slope in
$M_P$ is allowed, and we have obtained $A'$ compatible with 1 in most
cases.  All of this confirms 
that heavy quark symmetry is very well 
satisfied here. Moreover the
insensitivity of the results to the different modelling functions
gives us confidence, not only in the interpolation to the charm mass,
but also in the long extrapolation to the beauty mass. We stress, once
more, the total agreement between SS and SL results, both for the
final numbers and the features of the extrapolations.  Two examples of
the extrapolation to the heavy scale, corresponding to the $\Lambda$
and the $\Sigma$ are shown in Figs.~\ref{heavy}.

\section{Calculation of mass splittings}

Once the value of $\kappa_h$ is fixed to correspond to the physical
quark mass by matching $M_P$ to either the $D$- or the $B$-meson mass,
no large uncertainties are expected to occur in the measurement of
other charm or beauty hadron masses, which are largely determined by
that of the heavy quark. On the other hand, splittings in the masses
arise from the dynamics of the light quarks and their interactions
with the heavy quark. Their study provides a test of HQET as well as
important information on the size of various systematic effects.

The mass splittings are small quantities in comparison to the baryon
masses themselves. Thus, they are affected by relatively
larger statistical errors, as well as  being more sensitive to the
fitting and extrapolation procedures adopted. Therefore, a particularly
careful analysis is required. Once more, we will quote results obtained
by  fitting SS correlators; the SL correlators give consistent results,
although the statistical errors are appreciably larger.

\subsection{$\Lambda-$Pseudoscalar Meson Mass Splitting}
\begin{figure}
\begin{picture}(80,80)
\put(10,-7){\ewxy{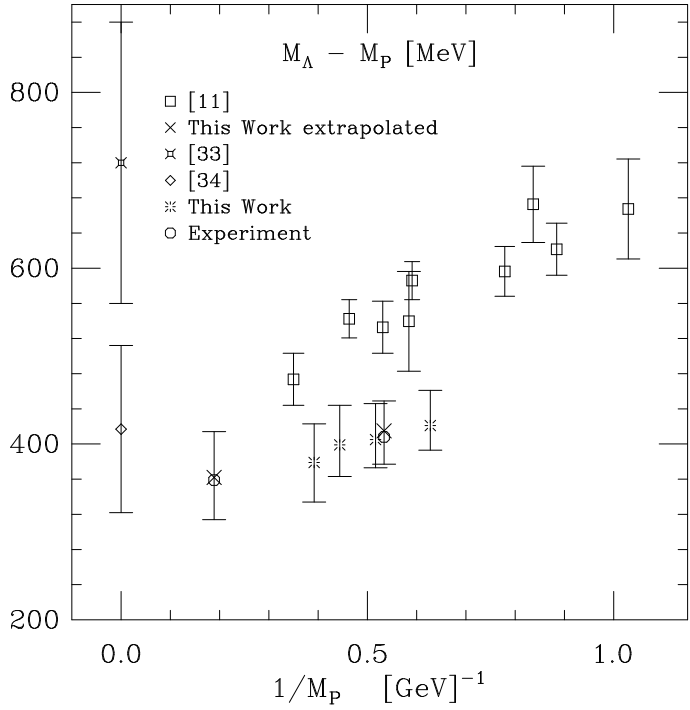}{130mm}}
\end{picture}
\caption{\em 
$M_{\Lambda} - M_P$ splitting: comparison of the values obtained by different
groups and using different fermion actions. The estimates are also compared with 
the experimental numbers.}
\vspace{-.5cm}
\label{figLAMBDAB}
\end{figure}

The $\Lambda-\mbox{ pseudoscalar meson}$ mass splitting is very 
precisely measured experimentally, expecially in the charm sector.
Therefore, as
we use the pseudoscalar mass to fix the value of the heavy quark
$\kappa$, the agreement of the lattice value of $M_{\Lambda}-M_{P}$
with experiment reflects the extent to which our
calculation properly incorporates the dynamics of the light degrees of
freedom. The amount of computational effort devoted to this 
calculation, both with static~\cite{bochi}-\cite{impre} and propagating Wilson
fermions~\cite{WUPPERTAL} is testimony to its importance. We summarize
the results obtained so far \cite{ALISTAIR,impre} and compare them with
the numbers from this study in Fig.~\ref{figLAMBDAB}.

For the analysis of this splitting we need the correlation function of
the pseudoscalar meson, which was determined in an earlier
simulation~\cite{fd} using the same heavy quarks as this study, but
with one additional light quark, corresponding to $\kappa_l=0.14262$.
We find that there is very little to be gained by determining the
difference of the masses from the time evolution of the ratio of the
$\Lambda$ and the pseudoscalar meson correlators\footnote{Since the
behaviour of the baryon and meson time slice correlators is different
close to the centre of the lattice, the ratio method is only safe if
one excludes the last few timeslices from the fitting range.}, as
opposed to obtaining it by subtracting the two masses determined
separately.  One reason is that the statistical errors associated with
our measurement of the $\Lambda$ mass are much larger than those of
the pseudoscalar, and would dominate the final uncertainty on the
splitting in any procedure.  Moreover, when the pseudoscalar
correlators were computed, only the heavy quark propagators were
smeared. Here the light quark propagators are also smeared, and
consequently the pseudoscalar and baryon correlation functions suffer
from different statistical fluctuations.  Therefore, we measure the
splitting by taking the difference of the baryon mass, obtained as
described in the previous section, and the meson mass, fitted in the
range $t\in[12,21]$, as in ref.~\cite{fd}.  The results, for all the
kappa values, are reported in Table~\ref{laPS}.

We perform a chiral extrapolation of the mass differences at each heavy
$\kappa$ value. Although we have simulated only two values of the light quark 
mass we have computed baryon correlators corresponding to two degenerate and 
one non-degenerate case; this last set of data, however, cannot be matched
with the mesonic data and cannot be used in the chiral extrapolation.
Hence, the chiral extrapolation is modelled by a linear 
function of the two
degenerate light-quark points. Both our results for the $\Lambda$ extrapolation (see section
3) and the evidence reported in ref.~\cite{strange} for the meson, 
justify  this procedure.

\begin{table}
\begin{center}
\begin{tabular}{||c|l|l|l|l||}
\hline\hline
$\kappa_{l1}/\kappa_{l2} $ & $\kappa_h=0.121$ & $\kappa_h=0.125$ & $\kappa_h=0.129$ & $\kappa_h=0.133$ \\
\hline
 $ 0.14144/0.14144 $&$0.212\err{7}{8}$&$ 0.216\err{8}{6}$&$ 0.222\err{8}{5}$&$ 0.229\err{8}{4}$\\ \hline
 $ 0.14226/0.14226 $&$0.173\err{10}{10}$&$0.178\err{10}{8}$&$0.182\err{10}{7}$&$0.189\err{10}{6}$\\ \hline
  $chiral/chiral   $&$0.131\err{15}{15} $&$0.138\err{15}{12} $&$0.140\err{14}{11} $&$0.145\err{14}{10} $\\ \hline
\hline
\end{tabular}
\end{center}
\caption{\em $\Lambda$-pseudoscalar meson mass splitting, in lattice units, 
obtained from the  difference of the fitted masses.}
\label{laPS}
\end{table}

\begin{figure}
\begin{picture}(80,100)
\put(10,-7){\ewxy{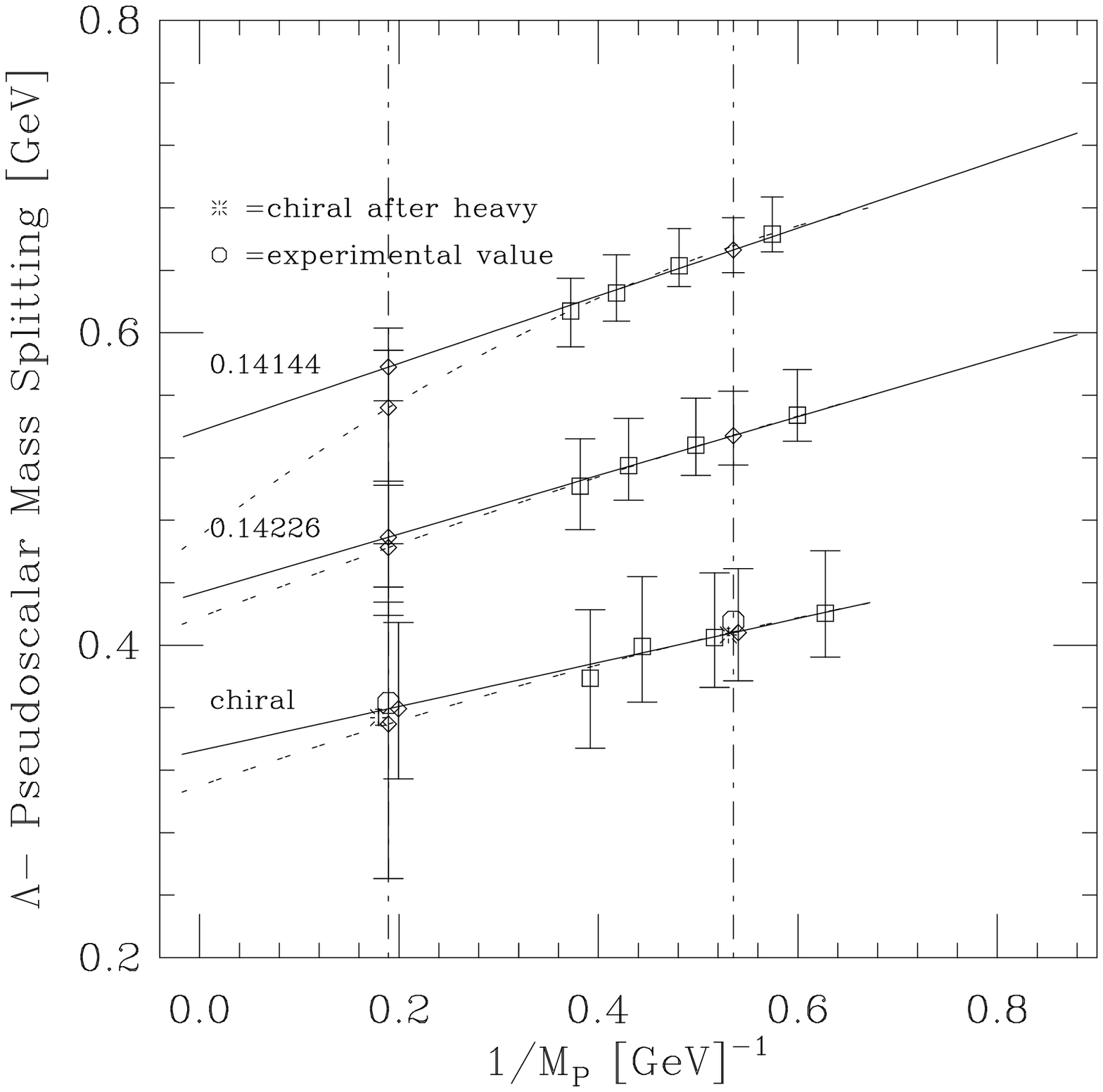}{130mm}}
\end{picture}
\caption{\em 
$M_{\Lambda}- M_P$ splitting  as obtained adopting the two procedures 
{\bf \rm A} and {\bf \rm B}. The solid lines correspond to the 
linear extrapolation in the inverse pseudoscalar mass and the dotted line is the 
same extrapolation modelled with a quadratic dependence. In the plot, the light 
$\kappa$ values are also indicated.
The results, which are consistent between both methods, are conpared
with the experimental values. The vertical dotted lines indicate 
$1/M_D$ and $1/M_B$.}
\label{figLAMBDAA}
\end{figure}

We performed the extrapolation to the physical pseudoscalar meson masses 
following two different procedures, in order to have a consistency
check ( see figure \ref{figLAMBDAA}).
\begin{itemize}
\item[{\bf A}] The splittings are first extrapolated  in the inverse heavy
quark mass,  according to the formula 
\[ 
\Big[M_{\Lambda}-M_{P}\Big](\kappa_h) = A + \frac{B}{M_P(\kappa_h)} +
\frac{C}{M^2_P(\kappa_h)},
\]
keeping the light quark mass fixed.
The linear and quadratic extrapolations  produce
indistinguishable numbers. Then the
two values of this splitting corresponding to the two degenerate
light-quark configurations are extrapolated to the chiral limit
for each heavy-quark mass.
\item[{\bf B}]
We employ the reverse procedure in which the mass splittings are first
extrapolated to the chiral limit, keeping the heavy quark mass fixed.
The results from the subsequent linear and quadratic extrapolations in
the heavy quark mass are once again compatible.
\end{itemize}
We conclude that the  behaviour of the $\Lambda$-pseudoscalar splitting is 
well represented by a linear function of both the light quark mass and the
inverse heavy quark mass. The results in physical units that 
we quote in Table~\ref{laPS2} are obtained under this assumption.

\begin{table}
\begin{center}
\begin{tabular}{||l|c|c||c||}
\hline\hline
      & procedure A [MeV]  &  procedure B [MeV] & experiment [MeV] \\
\hline
charm  &$  406\err{44}{29} $&$408\err{41}{31} $&$ 415(1) $  \\
beauty &$ 354\err{55}{46}$&$  359\err{55}{45} $&$ 362(50) $ \\
\hline\hline
\end{tabular}
\end{center}
\caption{\em Results for the $\Lambda-\mbox{ pseudoscalar}$ splitting, 
at the physical masses, corresponding to $a^{-1}=2.9\,GeV$.  The two
methods illustrated in the text produce essentially identical results,
in excellent agreement with the experimental number, given in the
last column.}
\label{laPS2}
\end{table}

\subsection{$\Sigma-\Lambda$ Mass Splitting}

We have obtained the $\Sigma-\Lambda$ mass splitting for the various $\kappa$ combinations, 
both by taking the difference of the fitted masses and by fitting the
time evolution  of the ratio of $\Sigma$ and $\Lambda$ correlators. 
The numbers obtained with the two methods are in good
agreement, but the second procedure yields appreciably smaller errors
and is smoother in  the chiral limit, proving that 
it is particularly appropriate when one compares two correlators
with a similar structure. The results at each value of the 
computed masses and
after the chiral extrapolation are given in
Table~\ref{tabSIGMA}.

\begin{figure}
\begin{picture}(80,80)
\put(10,-7){\ewxy{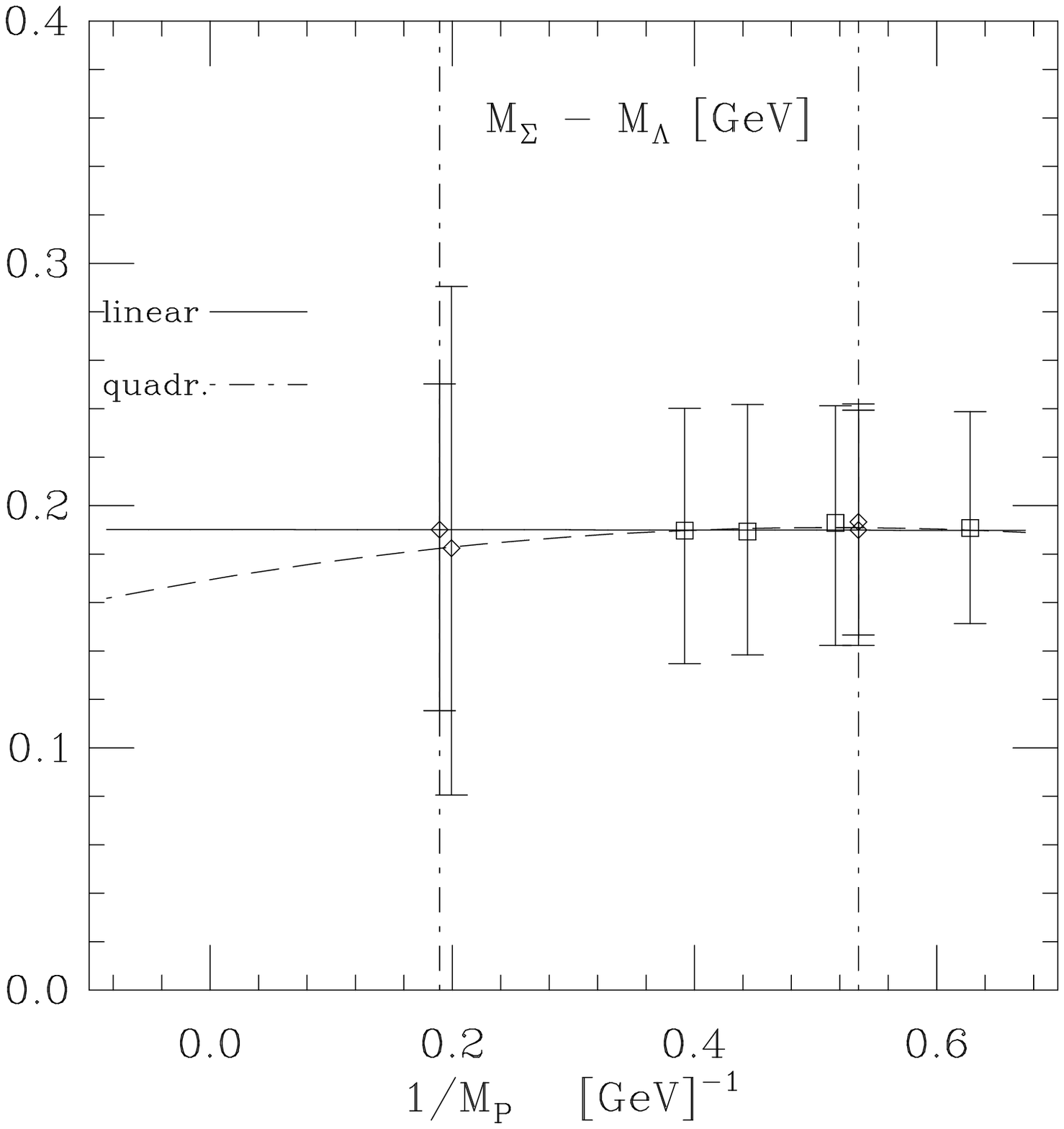}{130mm}}
\end{picture}
\caption{\em \protect$M_{\Sigma}-M_{\Lambda}\protect$ as a function of
\protect$1/M_P\protect$.
The linear and the quadratic extrapolation are shown.
The vertical dotted lines indicate $1/M_D$ and $1/M_B$,
rspectively.}
\vspace{-.5cm}
\label{figSIGMA}
\end{figure}

The dependence of the splitting on the heavy quark  mass
is extremely weak, suggesting that the $1/m_h$ corrections to the masses of the 
two baryons must be very similar and nearly cancel in the difference.
This feature makes it particularly simple to perform
the extrapolation to the physical masses, as the fits to linear and 
quadratic functions of the inverse pseudoscalar meson mass are
essentially indistinguishable. This is clearly visible in
Fig.~\ref{figSIGMA}.
We note that our results, presented in Table~\ref{tabSp1}, compare very well 
with experiment.

\begin{table}
\begin{center}
\begin{tabular}{||c|l|l|l|l||}
\hline\hline
$\kappa_{l1}/\kappa_{l2} $ & $\kappa_h=0.121$ & $\kappa_h=0.125$ & $\kappa_h=0.129$ & $\kappa_h=0.133$ \\
\hline
$0.14144/0.14144$&$\!( 0.37\err{6}{4})\!\times 10^{-1} $&$\!( 0.36\err{7}{4})\!\times 10^{-1} $&$\!(0.37\err{7}{4})\!\times 10^{-1} $&$\!(0.37\err{6}{4})\!\times 10^{-1} $ \\ \hline
$0.14144/0.14226$&$\!( 0.44\err{7}{8})\!\times 10^{-1} $&$\!( 0.44\err{8}{7})\!\times 10^{-1} $&$\!( 0.43\err{8}{7})\!\times 10^{-1} $&$\!(  0.44\err{8}{6})\!\times 10^{-1}$ \\ \hline
$0.14226/0.14226$&$\!( 0.50\err{12}{11})\!\times 10^{-1} $&$\!( 0.49\err{12}{10})\!\times 10^{-1} $&$\!( 0.52\err{10}{11})\!\times 10^{-1} $&$\!( 0.51\err{11}{8})\!\times 10^{-1} $ \\ \hline
$chiral/strange $& $\!( 0.54\err{15}{15})\!\times 10^{-1} $&$\!( 0.50\err{17}{13})\!\times 10^{-1} $&$\!( 0.59\err{13}{15})\!\times 10^{-1} $&$\!(0.57\err{14}{10})\!\times 10^{-1} $ \\ \hline
$chiral/chiral   $& $\!( 0.65\err{17}{18})\!\times 10^{-1} $&$\!( 0.65\err{18}{18})\!\times 10^{-1} $&$\!( 0.67\err{17}{17})\!\times 10^{-1} $&$\!(0.66\err{17}{14})\!\times 10^{-1} $\\
\hline\hline
\end{tabular}
\end{center}
\caption{\em Estimates of $M_{\Sigma}-M_{\Lambda}$ in lattice units, for the various kappa 
combinations, obtained with the ratio method. The extrapolations are 
linear, as they were for the masses themselves.}
\label{tabSIGMA}
\end{table}

\begin{table}[t]
\begin{center}
\begin{tabular}{||c|cc|cc||}\hline\hline
           & $\ \ h= $ & $\ charm $ & $\ \ h= $ & $\ beauty $ \\
           &   Exp.      & Latt.           &  Exp.       & Latt.      \\
\hline 
$\Lambda_{h}- P   $& [18]$\ \ 417(1)$ & $ 408\err{41}{31}\err{34}{33} $&[18]$\ \ 362(50) $&$359\err{55}{45}\err{27}{26} $\\\hline
\hline 
$\Sigma_{h}- \Lambda_{h} $& [18]$\ \ 169(2)$ & $ 190\err{50}{43}\err{13}{13} $&[4]$\ \ 173(11) $&$190\err{60}{75}\err{13}{13}$\\
$\Xi'_{h}- \Xi_{h}       $& [12]$\ \ \ \ \ \ \ 92$ & $ 166\err{40}{35}\err{12}{13} $&   &   $157\err{52}{64}\err{11}{11}$\\
\hline
\hline 
$\Sigma^*_{h}- \Sigma_{h}   $& [19]$\ \ \   77(6)$ & $ -17\err{12}{31}\er{3}{2} $&[4]$\ \ \ \ 56(16)$&$-6\err{4}{11}\er{1}{1} $\\
$\Xi^*_{h}- \Xi'_{h}       $&[12]$\ \ \ \ \ \ \ 83 
$ & $ -20\err{12}{24}\er{2}{3} $& &$-7\er{4}{8}\er{1}{1} $\\
$\Omega^*_{h}- \Omega'_{h}  $&  & $ -23\err{6}{14}\er{3}{2} $& &$ -8\er{2}{5}\er{1}{1} $\\
\hline\hline
\end{tabular}
\end{center}
\caption{\em Baryon-pseudoscalar meson and baryon-baryon mass splittings in MeV.
The available experimental data are also shown, together with the
corresponding references. The experimental errors on the $\Xi'_{c}-
\Xi_{c} $ and $\Xi^*_{c}- \Xi'_{c}$ splittings are not published. }
\label{tabSp1}
\end{table}


\subsection{Spin Splitting}

In the  HQET, the mass difference within the spin
doublets  $(\Sigma, \Sigma^*)$, $(\Xi', \Xi^*)$ and $(\Omega,
\Omega^*)$  is due to the coupling of the chromomagnetic moment of the heavy 
quark to the light degrees of freedom. It is therefore suppressed by inverse
powers of the heavy-quark mass and vanishes in the infinite mass limit.

Because the splitting is such a small number, it is difficult to
measure from our data using either the ratio of the correlators or the
 difference of the fitted masses.  We find that the difference of the
masses gives smoother chiral behaviour, and so we obtain our estimates
using this method.

\begin{figure}
\begin{picture}(80,100)
\put(10,-7){\ewxy{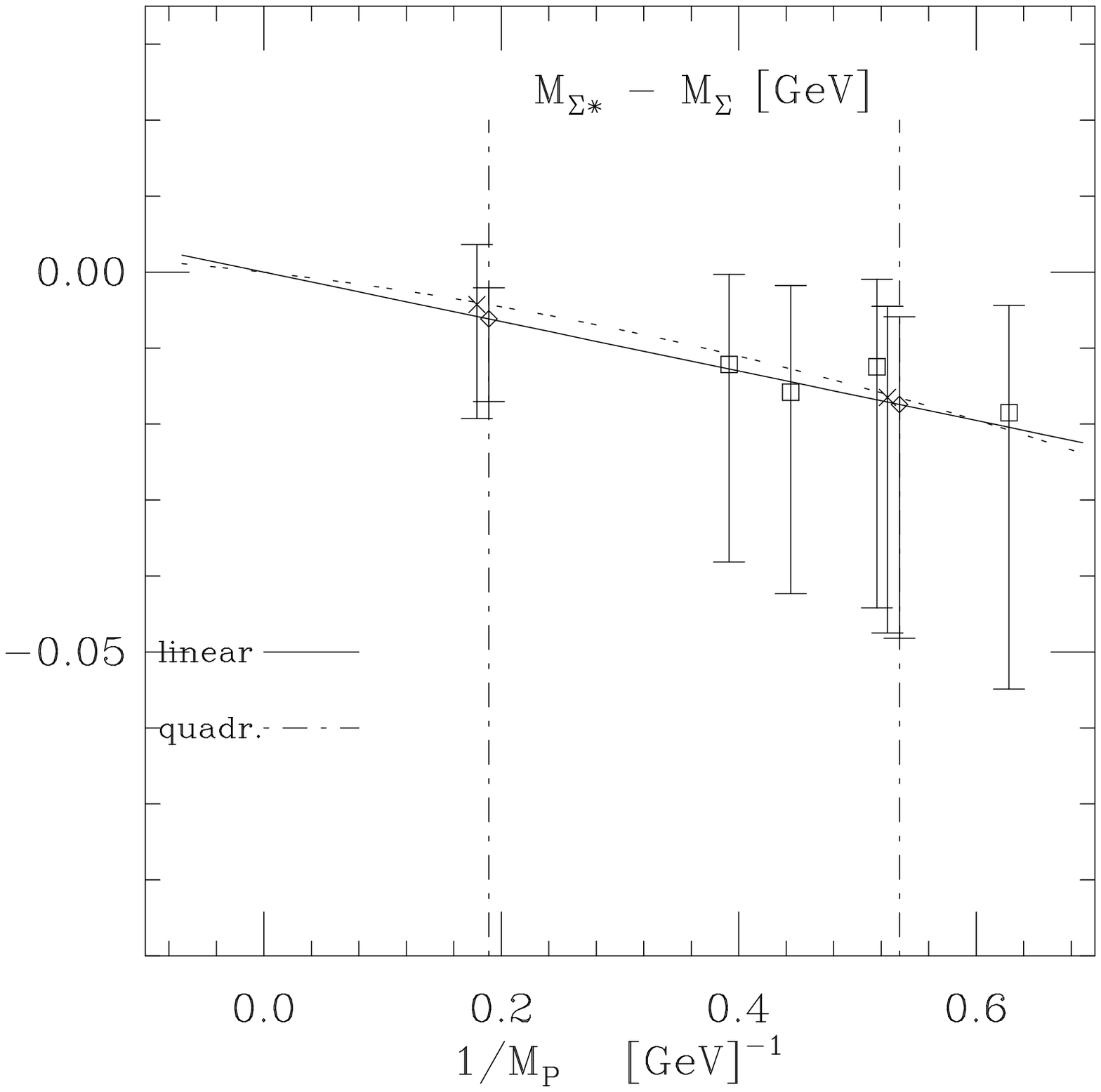}{130mm}}
\end{picture}
\caption{
\em \protect$M_{\Sigma^*} - M_{\Sigma}\protect$ splitting computed from
the mass difference, together with linear and quadratic
extrapolations. The vertical dotted lines indicate $1/M_D$ and $1/M_B$.}
\vspace{-.5cm}
\label{figSIGMA*}
\end{figure}

We present our measurements of the splittings for each $\kappa$
combination in Table~\ref{tabSPIN} and the extrapolated values at the
physical  masses in Table~\ref{tabSp1}. They are negative within
two standard deviations at fixed light-quark mass, but being
affected by large statistical errors, become compatible
with zero in the chiral limit. 
We obtain the splittings at the physical masses, extrapolating in the inverse 
heavy-quark mass, according to the following function
\be
\Big[M_{\Sigma^*}-M_{\Sigma}\Big](\kappa_h) = \frac{A}{M_P(\kappa_h)}
+ \frac{B}{M_P^2(\kappa_h)}.
\label{splitEQ}
\ee
The splitting has been constrained to vanish in the infinite heavy-quark 
mass limit, as predicted by the HQET. The two extrapolations obtained
either by including the quadratic term or by setting $B=0$ gave essentially 
indistinguishable results, as can be seen from Fig.~\ref{figSIGMA*},
and good $\chi^2$.
We also checked the consistency of our data with the predicted behaviour,
by adding a constant to the function (\ref{splitEQ}), i.e. by
allowing the spin-splitting to have a non-zero intercept, 
at $1/M_P=0$. We find that the values of the intercept are always compatible
with zero, being of the order -30 MeV, with errors of about one hundred.

Once more, the results were perfectly consistent. 
The values presented in Table~\ref{tabSp1} correspond to Eq.~(\ref{splitEQ}),
with $B=0$ since this parametrization fits the lattice data very well
and makes full use of heavy-quark scaling relations.

We note that our value for the $\Sigma_c^*-\Sigma_c$ splitting is
inconsistent with experiment, and in most cases our measured values are
consistent with zero.  This feature resembles the well-known
puzzle of the spin splitting in the mesonic sector \cite{MSCLOVER}, whose resolution is believed to lie in a combination
of discretisation and quenching effects, as argued in ref.~\cite{FERMILAB}.
In the light of this, and also considering the very large statistical errors,
firm conclusion about the consistency of
our results with heavy-quark scaling laws cannot be drawn.

\begin{table}
\begin{center}
\begin{tabular}{||c|l|l|l|l||}
\hline\hline
$\kappa_{l1}/\kappa_{l2} $ & $\kappa_h=0.121$ & $\kappa_h=0.125$ & 
$\kappa_h=0.129$ & $\kappa_h=0.133$ \\
\hline
 $\!\! 0.14144/0.14144\!\! $&$  \!(-0.53\err{17}{28})\!\times\! 10^{-2}\! $&$    \!(-0.60\err{18}{30})\!\times\! 10^{-2}\! $&$   \!(-0.74\err{21}{34})\!\times\! 10^{-2}\! $&$    \!(-0.93\err{22}{41})\!\times\! 10^{-2}\! $ \\ \hline
 $\!\! 0.14144/0.14226\!\! $&$ \!(-0.50\err{18}{39})\!\times\! 10^{-2}\! $&$   \!(-0.58\err{23}{42})\!\times\! 10^{-2}\! $&$     \!(-0.67\err{21}{49})\!\times\! 10^{-2}\! $&$    \!(-0.85\err{24}{58})\!\times\! 10^{-2}\! $ \\ \hline
 $\!\! 0.14226/0.14226\!\! $&$ \!(-0.49\err{26}{57})\!\times\! 10^{-2}\! $&$   \!(-0.63\err{31}{62})\!\times\! 10^{-2}\! $&$    \!(-0.68\err{32}{68})\!\times\! 10^{-2}\! $&$    \!(-0.84\err{38}{78})\!\times\! 10^{-2}\! $ \\ \hline
 $\!\! strange/strange\!\! $&$\!(-0.50\err{19}{41})\!\times\! 10^{-2}\! $&$\!(-0.58\err{23}{43})\!\times\! 10^{-2}\! $&$\!(-0.66\err{22}{52})\!\times\! 10^{-2}\! $&$   \!(-0.86\err{26}{61})\!\times\! 10^{-2}\! $ \\
 $chiral/strange $&$\!(-0.40\err{36}{75})\!\times\! 10^{-2}\! $&$\!(-0.50\err{45}{85})\!\times\! 10^{-2}\! $&$\!(-0.63\err{48}{86})\!\times\! 10^{-2}\! $&$\!(-0.97\err{56}{94})\!\times\! 10^{-2}\! $\\  
 $chiral/chiral   $&$\!(-0.42\err{41}{90})\!\times\! 10^{-2}\! $&$\!(-0.55\err{48}{92})\!\times\! 10^{-2}\! $&$\!(-0.43\errr{40}{110})\!\times\! 10^{-2}\! $&$  \!(-0.64\errr{49}{126})\!\times\! 10^{-2}\! $ \\ 
\hline\hline
\end{tabular}
\end{center}
\caption{\em Estimates of $M_{\Sigma^*}-M_{\Sigma}$ in lattice units at the various masses, 
obtained by taking the difference of the fitted masses, in
lattice units. 
The extrapolations are linear, following the same procedure adopted for the other splittings.}
\label{tabSPIN}
\end{table}

\section{Physical Results}

In this section, we present a summary of the results obtained in this
study, in a form which is easily comparable with the experimental data.
All masses are given with an asymmetric statistical error arising from
the bootstrap analysis, and a systematic error due solely to the
uncertainty in the scale (see Eq.~(\ref{eq:ainv})).

In Table~\ref{tab9} we quote the charm and beauty baryon masses,
together with the experimental values, where available. Our results
agree well with the experimental data for the charm sector, and also
for $\Lambda_b$ and $\Sigma_b$, despite the long extrapolation in the
heavy mass scale needed in these cases. This gives us confidence in
the reliability of our predictions for the masses of the undiscovered
charm and beauty baryons.  The quality of the results at the beauty
mass was certainly enhanced by the number of heavy quark masses
available for this investigation, which allowed us to try different
extrapolation procedures and to perform consistency checks. On the
other hand, we only have a limited sample of light quark masses.
Although the light extrapolations were always smooth and reasonable,
the chiral behaviour should be confirmed by using a larger number
of light quark masses.

We present the mass splittings of Section 4 in Table~\ref{tabSp1}.  In
those cases where comparison with experiment is possible, we also
compute the ratios of the splitting to the sum of the masses, to
eliminate most of the uncertainty in the scale.
These results were presented in the Introduction, see 
Eqns~(\ref{ratio1}-\ref{eq4}).  The residual systematic uncertainty, which is always
smaller than the statistical error, was not quoted.

In those cases where a meaningful comparison with experiment is
possible the agreement is very encouraging.  Unfortunately, the mass
differences, being small, are affected by large relative errors
varying between 10 and 30\%. Nevertheless, we stress the beautiful
agreement with the experimental data, both at the charm and at the
beauty mass. In particular, in our calculation of the
$\Lambda_h-\mbox{ pseudoscalar meson}$ splitting, the agreement with
experiment has significantly improved on previous
calculations, performed with the standard Wilson action.  We believe
that this success is further evidence of the advantages of using the
Sheikholeslami-Wohlert fermion action.


\section{Conclusions}

In this paper we have presented the result of a lattice study of heavy 
baryon spectroscopy. The spectrum of the eight lowest-lying heavy baryons,
containing a single heavy quark, can be computed using the three
baryonic operators in Eq.~(\ref{operators}). In addition to the
calculation of the $\Lambda$ and $\Xi$ masses
we have discussed how to
compute the spectrum of the spin  doublets, $(\Sigma,\Sigma^*)$,
$(\Xi',\Xi^*)$ and $(\Omega,\Omega^*)$, by isolating their contribution
to the correlation functions of the operators ${\cal O}_\mu$ and ${\cal
O}_\mu'$.

The computation of the mass spectrum proved feasible; the operators we have used have a good overlap
with the various baryon ground states, in part thanks to the smearing
both at the source and at the sink. Moreover, the
extrapolations in both the heavy and light quark masses are 
always smooth.
The agreement between our estimates of the baryon masses and the
experimental values is good, in both the charm and the beauty sectors.

The computed $\Lambda-\mbox{ pseudoscalar meson}$ mass splitting
is in  good agreement with experiment, in contrast with the results of
previous calculations performed with the Wilson fermion action. 
We believe that this is largely
due  to the use of the $O(a)$-improved action to remove systematic
effects. A similar positive conclusion can be drawn for the
$\Sigma-\Lambda$ splitting, although the statistical errors are
still of the order of $25-30\%$.

Our results are also in agreement with the predictions obtained with
other non-perturbative methods~\cite{POTENTIAL}-\cite{CHIRAL}, both
for the masses themselves and for the $\Sigma-\Lambda$ splitting.  In
the case of the $\Xi'-\Xi$ splitting, which is of the same nature as
the $\Sigma-\Lambda$ splitting, the calculation is complicated by the
mixing arising between the two particles, whose total quantum numbers
are the same. It has been noted \cite{CHIRAL} that such a mixing,
negligible in the heavy quark limit, would have the effect of
increasing the splitting.  Both our prediction and that of
Savage~\cite{CHIRAL} are higher than experiment~\cite{LAST1}. The
disagreement would hence get even worse if we were to take the mixing
into account.  We stress, anyway, that this experimental result is
still to be confirmed.

In this exploratory study the masses have been determined with 
reasonable precison, but further studies are required to reduce
both the statistical and systematic errors. 
The results presented in this paper are very encouraging and it looks likely 
that it will also be possible to measure the baryonic matrix elements.

\paragraph{Acknowledgements}
The authors wish to thank
Robert Coquereaux, Oleg Ogievetsky, Jay Watson and Jonathan Flynn 
for hepful discussions.
This research was supported by the UK Science and Engineering Research
Council under grants GR/G~32779, by the Particle Physics and Astronomy
Research Council under grant  GR/J~21347, by 
the European Union under contract CHRX-CT92-0051, by the University of
Edinburgh and by Meiko Limited.  CTS (Senior Fellow), 
and DGR (Advanced Fellow) 
acknowledge the support of the Particle Physics and \nopagebreak
Astronomy Research Council.  NS thanks the University of Rome ``La Sapienza'' for financial support.
JN and PU acknowledge the European Union for 
for their support through the award of a Fellowship, 
contracts Nos. CHBICT920066 and CHBICT930877.
OO is supported by JNICT under grant BD/2714/93.
 

\begin{appendix}

\section{Effect of smearing on baryon correlation functions}
\label{app2}

In this appendix we propose a description of the effect of a
non-Lorentz invariant smearing on a non-scalar operator. Interpolating
operators are smeared to improve their overlap with the physical
states one wishes to create or annihilate.  We discuss the case of
2-point correlation functions, restricting ourselves to the case of
spinorial operators which have overlap with a single type of spin 1/2
particle, like ${\cal O}_5(x)$ defined in Eqn.~(\ref{operators}).
Numerical evidence for this effect is also presented.  It should be
noted that the breaking of the Lorentz symmetry manifests itself only
when one considers correlators at finite momentum, and has therefore
no relevance to the determination of the spectrum.

Let us write the general expression for a local baryonic operator,
\be
J_{\rho}(x)= \Big(\psi(x)\Gamma\psi(x)\Big) \psi_{\rho}(x)\label{local},
\label{J1}
\ee
and consider the case where it has overlap with spin 1/2 states,
like ${\cal O}_5$.
It can destroy one such state, according to the relation
\be
\langle 0| J_{\rho}(0) |\vec{p}, r\rangle = Z u_{\rho}^{(r)}(\vec{p}).
\label{J2}
\ee
In Eqs. (\ref{J1},\ref{J2}),
$\rho$ and $r$ are the spinorial and polarization indices, respectively,
the antisymmetric sum over colour  
is understood and $\Gamma$ is a 
suitable combination of gamma and charge conjugation matrices. Finally,
the amplitude $Z$ is a Lorentz scalar.

In general, a smeared baryonic operator can be written as
\be
J^s_{\rho}(\vec{x},t)=\sum_{\vec{y},\vec{z},\vec{w}} 
f(|\vec{y}-\vec{x}|) f(|\vec{z}-\vec{x}|) f(|\vec{w}-\vec{x}|) 
\Big(\psi(\vec{y},t)\Gamma\psi(\vec{z},t)\Big) \psi_{\rho}(\vec{w},t).
\ee
Because the smearing is performed only in the spatial directions, 
Lorentz symmetry is lost and 
only spatial translations, rotations, parity and time reversal survive.
Therefore, the overlap
of the operator $J^s_{\rho}(\vec{x},t)$ with the state 
$| \vec{p}, r\rangle $
is given by the more general expression
\be
\langle 0| J_{\rho}^{s}(0) |\vec{p}, r \rangle = \left[
Z_1(|\vec{p}|) u^{(r)}(\vec{p}) + Z_2(|\vec{p}|) 
\gamma_0 u^{(r)}(\vec{p})\right]_{\rho}
\label{J2s}
\ee
where the amplitudes $Z_1$ and $Z_2$ may depend on the magnitude of
the three-momentum of the state $|\vec{p},r \rangle$, in accord
with the restricted symmetries of the system.

Let us consider the case of a SS two-point
correlator, for large $t$
\bea
G_{\rho\sigma}^{ss}(t,\vec{p})&=&\sum_{\vec{x}}
\langle 0| T[J^s_{\rho}(\vec{x},t) \bar{J}^s_{\sigma}(\vec{0},0)]|0\rangle 
e^{-i\vec{p}\cdot\vec{x}} \nn \\
&=& \sum_{|\vec{q},r\rangle} \sum_{\vec{x}} 
\frac{m}{E(\vec{p})}
\langle 0| J^s_{\rho}(\vec{0},0)| \vec{q},r\rangle
\langle \vec{q},r|\bar{J}^s_{\sigma}(\vec{0},0)|0\rangle e^{-i\vec{p}\cdot\vec{x}}
e^{i\vec{q}\cdot\vec{x}}  \\
&=& \sum_{r}\frac{m e^{-E(\vec{p})t}}{E(\vec{p})} 
\left[
\Big( Z_1(|\vec{p}|) + Z_2(|\vec{p}|)\gamma_0\Big)
u^{(r)}(\vec{p})\bar{u}^{(r)}(\vec{p})
\Big( Z_1(|\vec{p}|) + Z_2(|\vec{p}|)\gamma_0\Big) \right]_{\rho\sigma}.
\nn
\eea
This expression  can be conveniently rewritten as
\be
G^{ss}_{\rho\sigma}(t,\vec{p})=Z_s^2(|\vec{p}|)e^{-E(\vec{p})t} \left[
\frac{E+m-\alpha^2(E-m)}{4E}\one +\frac{E+m+\alpha^2(E-m)}{4E}\gamma_0 -
\frac{2\alpha}{4E}\vec{p}\cdot\vec{\gamma}\right]_{\rho\sigma}
\label{SS}
\ee
where $Z_s=Z_1+Z_2$, $\alpha=(Z_1-Z_2)/(Z_1+Z_2)$ 
and full Lorentz symmety is recovered when $\alpha=1$.
Eqn. (\ref{SS}) exhibits the following features:
\begin{itemize}  
\item
the exponential fall-off is not altered by the smearing. 
This was expected since a smearing function which only extends in the spatial 
directions preserves the form of the transfer matrix;
\item
it has the correct limit for 
$\alpha\to 1: \ \ G^{ss}(\vec{p},t) \propto G^{\rm loc}(\vec{p},t) $;
\item the degeneracy among the amplitudes of different 
spinorial components of the correlator is lifted;
\item
all the terms proportional to $\alpha$ vanish for 
$\vec{p}\to\vec{0}$,
so that the effect disappears at zero momentum\footnote{For this 
reason, this effect has no consequences for the results 
presented in this paper.}.
\end{itemize}

In the following, we will check this effect against the numerical data.
It is convenient to rewrite Eq.~(\ref{SS}) in the form
\be
G^{ss}(\vec{p},t) =\frac{e^{-E(\vec{p})t}}{2E} \left\{ mZ_m + \gamma_0 E Z_E
-\vec{p}\cdot\vec{\gamma}Z_p\right\}, \ \ \ {\rm with}
\left\{ \ba{lcl}
Z_E &=& \Big[ E+m +\alpha^2(E-m)\Big] \ds{\frac{Z_s^2}{2E}} \\
&&\\
Z_m &=& \Big[ E+m -\alpha^2(E-m)\Big] \ds{\frac{Z_s^2}{2m}} \\
&&\\
Z_p &=& \alpha Z_s^2 .
\ea \right.
\label{zetas}
\ee
We have found that $Z_E$ and $Z_m$ are compatible, as they should be,
considering that they differ by terms proportional to $\vec{p}^{ 2} /Em$,
which are very small for the values of momenta and masses in our simulation.
Furthermore $Z_p$ is significantly different from $Z_E$ and $Z_M$, which
shows that $\alpha$ is different from one.
$Z_s$ and $\alpha$ are given by
\be
Z^2_s =\frac{EZ_E+mZ_m}{E+m} \ \ \ \ 
{\rm and }\ \ \ \ \alpha=\frac{Z_p}{Z^2_s}.
\ee
The results of this exercise are presented in Table~\ref{app2tab1} 
for the $\Lambda$ baryon with momentum $\vec{p}=(\frac{2\pi}{L},0,0)$
and $(\frac{2\pi}{L},\frac{2\pi}{L},0)$, for masses corresponding to
the four values of $\kappa_h$ and $\kappa_{l1}=\kappa_{l2}=0.14144 $.
Using the estimated values of $Z_s^2$ and $\alpha$ we have also
recomputed  $Z_E$ and $Z_m$ (second row of Table \ref{app2tab1}) and
verified that they are  compatible with the fitted values.
The numerical results are  consistent with the picture illustrated
above, and the value of $\alpha$ is significantly different from 1,
demonstrating that such an effect  cannot be neglected. 

\begin{table}
\begin{center}

 Momentum $\vec{p}=(\ds{\frac{2\pi}{L}},0,0)$ 

\begin{tabular}{||l|c|c|c|c|c||}
\hline\hline
  & $ Z^2_E $ & $ Z^2_m $&$ Z^2_p $&$ Z_s^2 $&$ \alpha $ \\
\hline
 $\kappa_h =0.121 $ & $( 2.84\err{40}{37})\times 10^4$ & $(2.80\err{41}{36})\times 10^4  $ & $(1.86\err{17}{14})\times 10^4 $ & $(2.82\err{41}{37})\times 10^4 $ & $0.66\err{6}{7} $ \\ 
                    & $( 2.80\err{40}{36})\times 10^4$ & $(2.84\err{41}{37})\times 10^4  $ & & & \\ \hline
 $\kappa_h =0.125 $ & $( 2.86\err{43}{38})\times 10^4$ & $(2.79\err{44}{37})\times 10^4 $ & $(1.99\err{18}{15})\times 10^4$&$(2.83\err{44}{38})\times 10^4$&$0.71\err{7}{7} $ \\ 
                    & $(2.81\err{42}{37})\times 10^4 $ & $(2.85\err{44}{38})\times 10^4$ & & & \\ \hline
 $\kappa_h =0.129 $ & $( 2.83\err{35}{30})\times 10^4 $ & $(2.73\err{33}{29})\times 10^4 $ & $(2.17\err{20}{17})\times 10^4$&$(2.78\err{34}{30})\times 10^4$&$0.78\err{7}{7} $ \\ 
                    & $(2.76\err{34}{29})\times 10^4 $ & $(2.80\err{}{})\times 10^4 $ & & & \\ \hline
 $\kappa_h =0.133 $ & $( 2.77\err{35}{32})\times 10^4 $ & $(2.62\err{34}{29})\times 10^4 $ & $( 2.29\err{22}{17})\times 10^4$ & $(2.69\err{36}{30})\times 10^4 $ & $0.85\err{8}{7} $ \\ 
                    & $(2.68\err{35}{28})\times 10^4 $ & $(2.71\err{36}{32})\times 10^4 $ & & & \\ \hline
\hline 
\end{tabular}

\vspace{.5cm}

Momentum $\vec{p}=(\ds{\frac{2\pi}{L}},\ds{\frac{2\pi}{L}},0)$ 

\begin{tabular}{||l|c|c|c|c|c||}
\hline\hline
  & $ Z^2_E $ & $ Z^2_m $&$ Z^2_p $&$ Z_s^2 $&$ \alpha $ \\
\hline
 $\kappa_h =0.121 $ & $( 1.94\err{28}{35})\times 10^4$ & $(1.93\err{36}{28})\times 10^4 $ & $(1.09\err{14}{11})\times 10^4 $ & $(1.94\err{28}{36})\times 10^4$ & $0.56\err{11}{7} $ \\ 
                  & $(1.90\err{27}{35})\times 10^4$ & $(1.97\err{29}{37})\times 10^4 $ & & & \\ \hline
 $\kappa_h =0.125 $ & $( 1.93\err{32}{33})\times 10^4 $ & $(1.91\err{32}{33})\times 10^4 $ & $(1.16\err{15}{12})\times 10^4$ & $(1.92\err{32}{33})\times 10^4$ & $0.61\err{12}{8} $ \\ 
                    & $(1.89\err{31}{32})\times 10^4$ & $(1.96\err{32}{34})\times 10^4 $ & & & \\ \hline
 $\kappa_h =0.129 $ & $( 1.93\err{32}{33})\times 10^4$ & $(1.88\err{30}{33})\times 10^4 $ & $(1.25\err{16}{13})\times 10^4 $ & $(1.91\err{30}{34})\times 10^4$ & $0.66\err{14}{9} $ \\ 
                    & $(1.87\err{29}{31})\times 10^4$ & $(1.95\err{32}{35})\times 10^4$ & & & \\ \hline
 $\kappa_h =0.133 $ & $( 1.89\err{32}{34})\times 10^4$ & $(1.79\err{32}{32})\times 10^4 $ & $(1.34\err{18}{14})\times 10^4$ & $(1.84\err{32}{33})\times 10^4$ & $0.73\err{16}{10}$ \\ 
                    & $(1.80\err{31}{30})\times 10^4$ & $(1.88\err{34}{36})\times 10^4 $ & & & \\ \hline
\hline
\end{tabular}
\end{center}
\caption{\em
Estimates of $\alpha$ and $Z_s^2$ from the fitted values of 
$Z_m, Z_E, Z_p$.
In the second row corresponding to each 
$\kappa_h$, the fitted $Z_m$ and $Z_E$ are compared with the estimates using
Eq.~(\protect{\ref{zetas}}), and the measured values of $Z_S^2$ and $\alpha$.}
\label{app2tab1}
\end{table}

We conclude this appendix with a discussion of the SL correlators, 
whose spin structure is again different from that of local correlators.
This feature must be taken into account in the
analysis of three-point correlators when the inserted current operator is 
local.
Following the reasoning above, we find:
\newpage
\bea
G^{sl}(t,\vec{p})=Z_lZ_s(|\vec{p}|)e^{-E(\vec{p})t}\left\{
\frac{E+m-\alpha(E-m)}{4E}\one\right. &+&\frac{E+m+\alpha(E-m)}{4E}\gamma_0 \nn \\
&-&\left.
\frac{(1+\gamma_0)+\alpha(1-\gamma_0)}{4E}\vec{p}\cdot\vec{\gamma}\right\}.
\eea
As above, it is possible to measure $Z_sZ_l$ averaging 
$Z_E$ and $Z_m$. By doing so, we have extracted $Z_l$ for three different
values of the momentum, and the results are shown in
Table~\ref{app2tab2}.
The evidence that $Z_l$ is independent of $\vec{p}$ is a further
check of the validity of our interpretation.

\begin{table}
\begin{tabular}{||c||c|c|c|c||}
\hline\hline
Momentum & $ \kappa_h=0.121 $&$ \kappa_h=0.125 $&$ \kappa_h=0.129 $
&$ \kappa_h=0.133 $\\ 
\hline
$\vec{p}=\vec{0}$ & $(3.8\er{4}{3})\times 10^{-3}$&  $(3.6\er{3}{3})
\times 10^{-3}$& 
 $(3.5\er{3}{3})\times 10^{-3}$&  $(3.2\er{3}{2})\times 10^{-3}$\\ 
$\vec{p}=(\frac{2\pi}{L},0,0)$ &$(3.9\er{4}{4})\times 10^{-3}$& 
 $(3.7\er{4}{4})\times 10^{-3}$&  $(3.6\er{4}{3})\times 10^{-3}$&  
$(3.3\er{4}{3})\times 10^{-3}$\\ 
$\vec{p}=(\frac{2\pi}{L},\frac{2\pi}{L},0)$& $(3.8\er{5}{4})\times 10^{-3}$&  
$(3.7\er{5}{4})\times 10^{-3}$&  $(3.5\er{5}{4})\times 10^{-3}$&  
$(3.3\er{5}{4})\times 10^{-3}$\\
\hline
\hline
\end{tabular}
\caption{\em Values of $Z_l=(Z_lZ_s)/\protect\sqrt{Z_s^2}$, 
for the four heavy masses and $\kappa_{l1}=\kappa_{l2}=0.14144$. }
\label{app2tab2}
\end{table}
\end{appendix}


\def \ajp#1#2#3{Am. J. Phys. {\bf#1}, #2 (#3)}
\def \apny#1#2#3{Ann. Phys. (N.Y.) {\bf#1}, #2 (#3)}
\def \app#1#2#3{Acta Phys. Polonica {\bf#1}, #2 (#3)}
\def \arnps#1#2#3{Ann. Rev. Nucl. Part. Sci. {\bf#1}, #2 (#3)}
\def \cmts#1#2#3{Comments on Nucl. Part. Phys. {\bf#1}, #2 (#3)}
\def \cn{Collaboration}
\def \cp89{{\it CP Violation,} edited by C. Jarlskog (World Scientific,
Singapore, 1989)}
\def \efi{Enrico Fermi Institute Report No. EFI}
\def \f79{{\it Proceedings of the 1979 International Symposium on Lepton and
Photon Interactions at High Energies,} Fermilab, August 23-29, 1979, ed. by
T. B. W. Kirk and H. D. I. Abarbanel (Fermi National Accelerator Laboratory,
Batavia, IL, 1979}
\def \hb87{{\it Proceeding of the 1987 International Symposium on Lepton and
Photon Interactions at High Energies,} Hamburg, 1987, ed. by W. Bartel
and R. R\"uckl (Nucl. Phys. B, Proc. Suppl., vol. 3) (North-Holland,
Amsterdam, 1988)}
\def \ib{{\it ibid.}~}
\def \ibj#1#2#3{~{\bf#1}, #2 (#3)}
\def \ichep72{{\it Proceedings of the XVI International Conference on High
Energy Physics}, Chicago and Batavia, Illinois, Sept. 6 -- 13, 1972,
edited by J. D. Jackson, A. Roberts, and R. Donaldson (Fermilab, Batavia,
IL, 1972)}
\def \ijmpa#1#2#3{Int. J. Mod. Phys. A {\bf#1}, #2 (#3)}
\def \ite{{\it et al.}}
\def \jpb#1#2#3{J.~Phys.~B~{\bf#1}, #2 (#3)}
\def \lkl87{{\it Selected Topics in Electroweak Interactions} (Proceedings of
the Second Lake Louise Institute on New Frontiers in Particle Physics, 15 --
21 February, 1987), edited by J. M. Cameron \ite~(World Scientific, Singapore,
1987)}
\def \ky85{{\it Proceedings of the International Symposium on Lepton and
Photon Interactions at High Energy,} Kyoto, Aug.~19-24, 1985, edited by M.
Konuma and K. Takahashi (Kyoto Univ., Kyoto, 1985)}
\def \mpla#1#2#3{Mod. Phys. Lett. A {\bf#1}, #2 (#3)}
\def \nc#1#2#3{Nuovo Cim. {\bf#1}, #2 (#3)}
\def \np#1#2#3{Nucl. Phys. {\bf#1}, #2 (#3)}
\def \PDG{Particle Data Group, L. Montanet \ite, \prd{50}{1174}{1994}}
\def \pisma#1#2#3#4{Pis'ma Zh. Eksp. Teor. Fiz. {\bf#1}, #2 (#3) [JETP Lett.
{\bf#1}, #4 (#3)]}
\def \pl#1#2#3{Phys. Lett. {\bf#1}, #2 (#3)}
\def \pla#1#2#3{Phys. Lett. A {\bf#1}, #2 (#3)}
\def \plb#1#2#3{Phys. Lett. B {\bf#1}, #2 (#3)}
\def \pr#1#2#3{Phys. Rev. {\bf#1}, #2 (#3)}
\def \prc#1#2#3{Phys. Rev. C {\bf#1}, #2 (#3)}
\def \prd#1#2#3{Phys. Rev. D {\bf#1}, #2 (#3)}
\def \prl#1#2#3{Phys. Rev. Lett. {\bf#1}, #2 (#3)}
\def \prp#1#2#3{Phys. Rep. {\bf#1}, #2 (#3)}
\def \ptp#1#2#3{Prog. Theor. Phys. {\bf#1}, #2 (#3)}
\def \rmp#1#2#3{Rev. Mod. Phys. {\bf#1}, #2 (#3)}
\def \rp#1{~~~~~\ldots\ldots{\rm rp~}{#1}~~~~~}
\def \si90{25th International Conference on High Energy Physics, Singapore,
Aug. 2-8, 1990}
\def \slc87{{\it Proceedings of the Salt Lake City Meeting} (Division of
Particles and Fields, American Physical Society, Salt Lake City, Utah, 1987),
ed. by C. DeTar and J. S. Ball (World Scientific, Singapore, 1987)}
\def \slac89{{\it Proceedings of the XIVth International Symposium on
Lepton and Photon Interactions,} Stanford, California, 1989, edited by M.
Riordan (World Scientific, Singapore, 1990)}
\def \smass82{{\it Proceedings of the 1982 DPF Summer Study on Elementary
Particle Physics and Future Facilities}, Snowmass, Colorado, edited by R.
Donaldson, R. Gustafson, and F. Paige (World Scientific, Singapore, 1982)}
\def \smass90{{\it Research Directions for the Decade} (Proceedings of the
1990 Summer Study on High Energy Physics, June 25--July 13, Snowmass,
Colorado),
edited by E. L. Berger (World Scientific, Singapore, 1992)}
\def \tasi90{{\it Testing the Standard Model} (Proceedings of the 1990
Theoretical Advanced Study Institute in Elementary Particle Physics, Boulder,
Colorado, 3--27 June, 1990), edited by M. Cveti\v{c} and P. Langacker
(World Scientific, Singapore, 1991)}
\def \yaf#1#2#3#4{Yad. Fiz. {\bf#1}, #2 (#3) [Sov. J. Nucl. Phys. {\bf #1},
#4 (#3)]}
\def \zhetf#1#2#3#4#5#6{Zh. Eksp. Teor. Fiz. {\bf #1}, #2 (#3) [Sov. Phys. -
JETP {\bf #4}, #5 (#6)]}
\def \zpc#1#2#3{Zeit. Phys. C {\bf#1}, #2 (#3)}
\def \zpd#1#2#3{Zeit. Phys. D {\bf#1}, #2 (#3)}


\vspace{-5mm}
\begin{thebibliography}{99}

\bibitem{LEP1}
C.~Albajar {\it et al.}, Phys.~Lett.\ {\bf B273}, (1991) 540;\\
G.~Bari {\it et al.}, Nuovo Cimento {\bf A104} (1991) 1787. 

\bibitem{LEPSEMIL}
ALEPH Collaboration 
D.~Decamp \ite,\plb{278}{367}{1992}; \\
OPAL Collaboration 
P.~Acton \ite,\plb{281}{394}{1992}; \\
ALEPH Collaboration 
D.~Buskulic. \ite,\plb{294}{145}{1992}; \\
DELPHI Collaboration 
P.~Abreu \ite,\plb{311}{379}{1993}. 

\bibitem{Peer1}
S. F. Biagi \ite, \zpc{28}{175}{1985}; ARGUS \cn, H. Albrecht
\ite, \plb{288}{367}{1992}; \\
Fermilab E687 \cn, P. L. Frabetti \ite,
\plb{300}{190}{1993}; \ibj{338}{106}{1994}; contribution to LISHEP95, Rio de
Janeiro, February 1995; \\
  CERN WA89 \cn, M. I. Adamovich \ite,
Max-Planck-Institut report MPI-H-V27-1995, June, 1995, and references therein.

\bibitem{Peer2}
Several contributions at the EPS HEP-95 Conference focused on this topic.
DELPHI Coll., DELPHI-95-107, Contribution \# eps0565, Brussel, 1995.


\bibitem{NEUBERT}
For a review, see, e.g.,
M.~Neubert; Heavy Quark Symmetry. Phys. Rept. {\bf 245} (1994) 245.

\bibitem{DESY}
J.G.~K\"orner, D.~Pirjol and M.~Kr\"amer; Prog.~Part.~Nucl.~Phys.
{\bf 33} (1994) 787.

\bibitem{POTENTIAL} 
A.~De R\'ujula, H.~Georgi and S.L.~Glashow.
Phys.~Rev.\ {\bf D12} (1975) 147; \\
A.~Martin and J.M.~Richard; Phys.~Lett..\ {\bf B185} (1987) 426, \\
and references therein.

\bibitem{rosner}
J.L. Rosner, EFI-95-48, hep-lat/9508252; \\
R. Roncaglia, D.B. Lichtenberg and E. Predazzi.
IU/NTC 95-03  hep-ph/9502251.

\bibitem{CHIRAL}
M.J. Savage, Report-no: CMU-HEP95-11, hep-ph/9508268;
see also M.K.~Banerjee and J.~Milana.
Preprint DOE/ER/40762--051, UMPP \#95--058, hep-ph/9410398. 


\bibitem{lat94}
UKQCD Collaboration (N. Stella) ; 
Nucl.~Phys. ~B (Proc. ~Suppl.) {\bf 42} (1995) 367;\\
C.~Alexandrou {\it et al.}, 
Nucl.~Phys. B (Proc. Suppl.) {\bf 42} (1995) 297.

\bibitem{WUPPERTAL}
C.~Alexandrou \ite, \plb{337}{340}{1994}.


\bibitem{LAST1}WA89 Collaboration, P. Avery et al. CLNS 95/1352, CLEO 95-14, hep-ex/9508010.

\bibitem{Rosner1}
J.L. Rosner, Prog. Theor. Phys. {\bf 66} (1981) 1422.


\bibitem{MSCLOVER} UKQCD Collaboration C.R.~Allton {\it et al.}, 
\plb{292}{408}{1992}.

\bibitem{FERMILAB}
A.S.~Kronfeld, P.B.~Mackenzie. Ann.~Rev.~Nucl.~Part.~Sci. {\bf 43} (1993) 793,
{\em and references therein }.

\bibitem{LAST2}G.P. Lepage and P.B. Mackenzie, \prd{48}{2250}{1993}.\\
UKQCD Collaboration, work in progress.

\bibitem{PDB}
Particle Data Group, (L.\,Montanet et al.) \prd{50}{}{1994}.



\bibitem{SKAT}
SKAT Collaboration, V.V.~Ammosov et al., paper \#42 submitted to
XVI International Symposium on Lepton-Photon Intereactions, Cornell University,
Ithaca (1993)


\bibitem{BEN}
M.~Benmerrouche \ite, \prc{39}{2339}{1989}.

\bibitem{RS}
W.~Rarita and J.~Schwinger.
Phys.~Rev.\ {\bf 60}, (1941) 61.

\bibitem{LURIE}
D.~Lur\'\i e. Particles and Fields.
Interscience Publishers, (1968).


\bibitem{a_units}
UKQCD Collaboration, C.R.~Allton \ite, \np{B407}{331}{1993}.
\bibitem{SW} 
B.~Sheikholeslami and R.~Wohlert, \np{B259}{572}{1985}.

\bibitem{HEATLIE}
G.~Heatlie, C.T.~ Sachrajda, G.~Martinelli, C.~Pittori and G.C.~Rossi,
\np{B352}{266}{1991}.

\bibitem{strange}
 UKQCD Collaboration, C.R.~Allton \ite, \prd{49}{474}{1994}.

\bibitem{gupta} R.~Gupta, to appear in the Proceedings of
the Conference LATTICE95, Melbourne. hep-lat/9512021.


\bibitem{Sommer} R.~Sommer,\np{B411}{839}{1994};
G. de Divitiis \ite, \np{B437}{447}{1994} .

\bibitem{biele_how} UKQCD Collaboration (H.~Wittig); 
Nucl.~Phys. B (Proc. Suppl.) {\bf 42} (1995) 288.

\bibitem{jonivar} UKQCD Collaboration, D.S.~Henty \ite,
\prd{51}{5323}{1995}.

\bibitem{SMEARING}
UKQCD Collaboration, C.R.~Allton \ite, \prd{47}{5128}{1993}.


\bibitem{bochi}
M. Bochicchio, G. Martinelli, C.R. Allton, C.T. Sachrajda and D.B. Carpenter.
\np{B372}{403}{1992}.

\bibitem{ALISTAIR}
UKQCD Collaboration (A.K.~Ewing);
Nucl.~Phys. B (Prog. Suppl.) {\bf B42} (1995) 331.

\bibitem{impre} 
UKQCD Collaboration (A.K.~Ewing \ite); Southampton preprint SHEP 95-20. hep-lat/9508030

\bibitem{fd}
UKQCD Collaboration, R.M.~Baxter \ite, \prd{49}{1594}{1994}.


\bibitem{MANDULA}
J.E.~Mandula and E.~Shpiz, \np{B232}{180}{1984}.

\bibitem{Chen} L.C.~Chen and J.L.~Birman, J. Math. Phys. {\bf 12} (1971) 2454
\bibitem{Mandula} J.E.~Mandula, G.~Zweig and J.~Govaerts, \np{B228}{91}{1983}
\np{B228}{109}{1983}.

\end{thebibliography}
\end{document}